\documentclass[useAMS,usenatbib]{mnras}%,epsfig - inside[]

\usepackage{amsmath, amssymb}
\usepackage{epsfig}         % For including postscript figures;
\usepackage{graphicx, float}% For including postscript figures; pdf, jpg graphics
\usepackage{multirow}
\usepackage{tabularx}
\usepackage{bm}            % for bold mathematical symbols

\usepackage{natbib}
\usepackage{aas_macros}

\usepackage{url}
\usepackage[]{hyperref}
\usepackage[hyphenbreaks]{breakurl}
\newcolumntype{L}[1]{>{\raggedright\arraybackslash}p{#1}}
\newcolumntype{C}[1]{>{\centering\arraybackslash}p{#1}}
\newcolumntype{R}[1]{>{\raggedleft\arraybackslash}p{#1}}
\usepackage{xcolor,colortbl}
\definecolor{Gray}{gray}{0.85}
\newcolumntype{G}{>{\columncolor{Gray}}r}

\def\Mpch{~h^{-1} {\rm Mpc}}

\def\MpchVolumeInverse{~h^3 {\rm Mpc}^{-3}}

\def\Msolar{~h^{-1} \rm{M}_{\odot}}

\def\kms{~\rm{km/s}}

\newcommand{\MI}{\textsc{MS}}

\newcommand {\reff} {R_{\rm eff}}
\newcommand {\vrad} {v_{\parallel}}
\newcommand {\vradlin} {v_{\parallel,\rm\;lin}}
\newcommand {\grad} {g_{\parallel}}
\newcommand {\dist} {\mathcal{D}}
\newcommand {\distVec} {\bm{\mathcal{D}}}
\newcommand {\shear} {\gamma_{\rm t}}
\newcommand {\conv} {\kappa}

\newcommand {\tin} {t_{\rm in}}
\newcommand {\tout} {t_{\rm out}}

\newcommand {\boundary} {boundary}
\newcommand{\Vector}[1]{\mathbf{#1}}

\newcommand{\gsim}{\raisebox{-0.3ex}{\mbox{$\stackrel{>}{_\sim} \,$}}}
\newcommand{\lsim}{\raisebox{-0.3ex}{\mbox{$\stackrel{<}{_\sim} \,$}}}
\newcommand{\eq}[1]{Eq. \eqref{#1}}

\newcommand{\refsec}[1]{Sec. \ref{#1}}
\newcommand{\refsecs}[2]{Secs. \ref{#1}-\ref{#2}}

\newcommand{\refappendix}[1]{Appendix \ref{#1}}

\newcommand{\reffig}[1]{Fig. \ref{#1}}

\newcommand{\reffigS}[2]{Figs \ref{#1} and \ref{#2}}

\newcommand{\figDir}{fig_pdf/}

\newcommand{\SvdW}{\hyperlink{labelHypertarget}{SvdW}}

\usepackage{color}
\definecolor{mycolor}{rgb}{.0,.3,1.}

\newcommand{\MCn}[1]{#1} % adds NEW text
\newcommand{\MCd}[1]{} % DELETE text
\newcommand{\MCc}[1]{} % COMMENT text
\newcommand{\MCq}[1]{} % emphasize text for question

\newcommand{\MCnn}[1]{#1} % adds NEW text

\newcommand{\MCnnn}[1]{#1} % adds NEW text
\title[The boundary profile of voids]
{The view from the boundary: a new void stacking method}

\author[Cautun, Cai \& Frenk]
{\parbox{\textwidth}{
        Marius Cautun$^{1}$\thanks{E-mail : m.c.cautun@durham.ac.uk},
        Yan-Chuan Cai$^{2,1}$
        and Carlos S. Frenk$^{1}$
        \vspace{.0cm}} \\
$^1$   Department of Physics, Institute for Computational Cosmology, Durham University, South Road Durham DH1 3LE, UK \\
$^2$   Institute for Astronomy, University of Edinburgh, Royal Observatory, Edinburgh EH9 3HJ, UK \\
}

\begin{document}

% \date{Accepted .... Received ...; in original form ...}
% \pagerange{\pageref{firstpage}--\pageref{lastpage}} \pubyear{2012}

\maketitle
%\label{firstpage}

\begin{abstract}
    We introduce a new method for stacking voids and deriving their profile that greatly increases the potential of voids as a tool for precision cosmology. Given that voids are distinctly non-spherical and have most of their mass at their edge, voids are better described relative to their boundary rather than relative to their centre, as in the conventional spherical stacking approach. The \emph{\boundary{} profile} is obtained by computing the distance of each volume element from the void boundary. Voids can then be stacked and their profiles computed as a function of this boundary distance. This approach enhances the weak lensing signal of voids, both shear and convergence, by a factor of two when compared to the spherical stacking method. It also results in steeper void density profiles that are characterised by a very slow rise inside the void and a pronounced density ridge at the void boundary. The resulting \boundary{} density profile is self-similar when rescaled by the thickness of the density ridge, implying that the average rescaled profile is independent of void size. The \boundary{} velocity profile is characterized by outflows in the inner regions whose amplitude scales with void size, and by a strong inflow into the filaments and walls delimiting the void. This new picture enables a straightforward discrimination between collapsing and expanding voids both for individual objects as well as for stacked samples.
\end{abstract}
%, in qualitative agreement with theoretical models of expanding spherical underdensities.


%%%%%%%%%%%%%%%%%%%%%%%%%%%%%%%%%%%%%%%%%%%%%%%%%%%%%%%%%%%%%%%%%%%%%%%%%%%%%
\section{Introduction}
\label{sec:introduction}
Cosmic voids represent a potentially powerful tool for measuring the cosmological parameters and probing the nature of dark energy \citep[e.g.][]{Li2011,Bos2012,Lavaux2012,Sutter2012,Cai2014a,Cai2014b,Hamaus2014c}. Most cosmological constraints are derived from the structure and dynamics of voids, which are a probe of modified gravity models \citep{Li2012,Clampitt2013,Cai2014,Barreira2015} as well as of the nature of dark matter \citep{Hellwing2010,Yang2014,Massara2015}. In the former case, void profiles are sensitive to the presence of a fifth force, which, while screened in higher density regions, can be large in voids. Such a force leads to emptier and larger voids, due to the faster evacuation of matter from low density regions \citep{Peebles2010,Clampitt2013}. In the latter case, replacing cold dark matter by warm dark matter or including massive neutrinos would lead to less evolved voids, and hence to shallower density profiles.
%
%Extracting the cosmological information

Up to now, the density and velocity structure of voids has been studied through the use of spherical profiles motivated by the fact that stacking many voids results into spherically symmetric structures \citep[e.g.][]{vandeWeygaert1993,Padilla2005,Ceccarelli2013,Ricciardelli2013,Hamaus2014b,Nadathur2015a}. But individual voids are distinctly non-spherical. While the simple picture of an expanding underdensity in a uniform background suggests that voids should become more spherical as they evolve \citep[][]{Icke1984}, in reality, voids are not isolated and this simplified picture does not hold. There are two major factors that affect the evolution of voids. Firstly, contrary to the case of collapsed structures, void evolution is strongly affected by the tidal field of the surrounding distribution of matter \citep{Platen2008,vandeWeygaert2011}. Secondly, as voids expand, they are squeezed by neighbouring voids. These effects lead to present-day voids that have highly complex shapes \citep{Platen2007,Platen2008,Neyrinck2008,Sutter2012a,Nadathur2014}. 

The diversity of void shapes makes the traditional stacking procedure suboptimal for extracting cosmological information. Simply put, the cosmological constraints are derived by comparing the density inside voids with that at their boundaries. For example, in some modified theories of gravity the inner regions of voids are emptier than in the standard cosmological model, with the evacuated matter deposited at the void boundaries. Stacking randomly oriented voids of various shapes leads to an overlap of the voids inner regions and boundaries. This ``blurring'' decreases the density contrast between the inner and outer parts of voids, leading to a lower signal. \MCn{In addition, there is ambiguity in the definition of the void centre used for spherical stacking, with different choices resulting in different density profiles \citep[e.g.][]{Nadathur2015b}.}

In this work, we introduce a new method of both measuring void profiles and stacking voids by taking into account their shape. In contrast to the spherical method, we propose that void profiles should be measured with respect to the void boundary. This leads to a much sharper distinction between the inside, boundary and outside of voids, resulting in at least two major gains. Firstly, it leads to a better understanding of the structure and dynamics of cosmic voids enabling a closer comparison with analytical theories of void evolution. Secondly, it increases the stacked lensing signal of voids, which is the best probe for measuring void density profiles \citep{Higuchi2013,Krause2013}.

The paper is organized as follows. In \refsec{sec:whats_about} we outline the new method by applying it to a simplified void model; in \refsec{sec:data} we describe the cosmological simulation to which we apply the method as well as the void catalogues we construct from it; in \refsecs{sec:density_profile}{sec:weak_lensing} we present the density, velocity and weak lensing profiles obtained using the new \boundary{} stacking approach. We conclude with a short discussion and summary in \refsec{sec:conclusions}.

%%%%%%%%%%%%%%%%%%%%%%%%%%%%%%%%%%%%%%%%%%%%%%%%%%%%%%%%%%%%%%%%%%%%%%%%%%%%%%
%\section{A shape independent void profile}
\section{The boundary profile of voids}
\label{sec:whats_about}

\begin{figure}
     \centering
     \includegraphics[width=0.66\linewidth,angle=0]{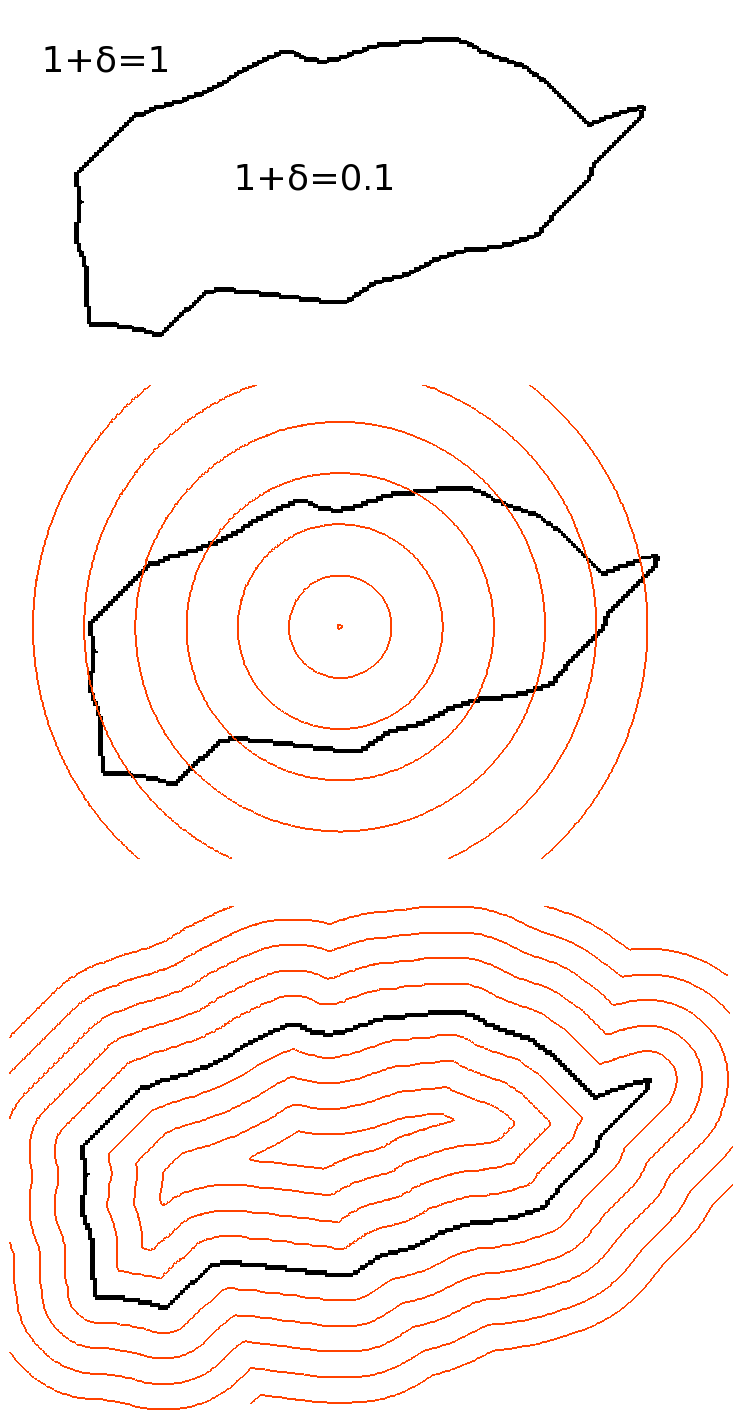}
     \caption{ Illustration of the new method for measuring void profiles. The top panel shows the void boundary, with the actual void shape selected randomly from voids found in an N-body simulation. For simplicity, the void is assigned a constant density, $1+\delta_{\rmn{inside}}=0.1$, inside its boundaries and is embedded in a uniform background with $1+\delta_{\rmn{outside}}=1$. The mass evacuated from inside the void is deposited at the void boundary, which has $1+\delta_{\rmn{boundary}}=30$. The center panel shows the spherical shells around the barycentre of the void that are used for computing the spherical profile. The bottom panel shows lines of equal distance from the void boundary (thick black curve) that are used for computing the \boundary{} profile proposed in this paper. }
     \label{fig:toy_example}
\end{figure}

Here we give an overview of the proposed method for computing \boundary{} void profiles, which we illustrate using a simplified model of a void. We construct a void by randomly selecting a shape for it from a cosmological N-body simulation. A cross section through the boundary of the void is shown in the top panel of \reffig{fig:toy_example}. For simplicity, the inner region of the void is assigned constant density, $1+\delta_{\rmn{inside}}=0.1$, where, $\delta=\tfrac{\rho}{\overline{\rho}}-1$, denotes the density contrast. The void is embedded within a uniform background, $1+\delta_{\rmn{outside}}=1$, and the mass evacuated from within the void is deposited uniformly on the boundary, which is shown as a solid curve. 

\begin{figure}
     \centering
    $\begin{array}{c}
        \includegraphics[width=.93\linewidth,angle=0]{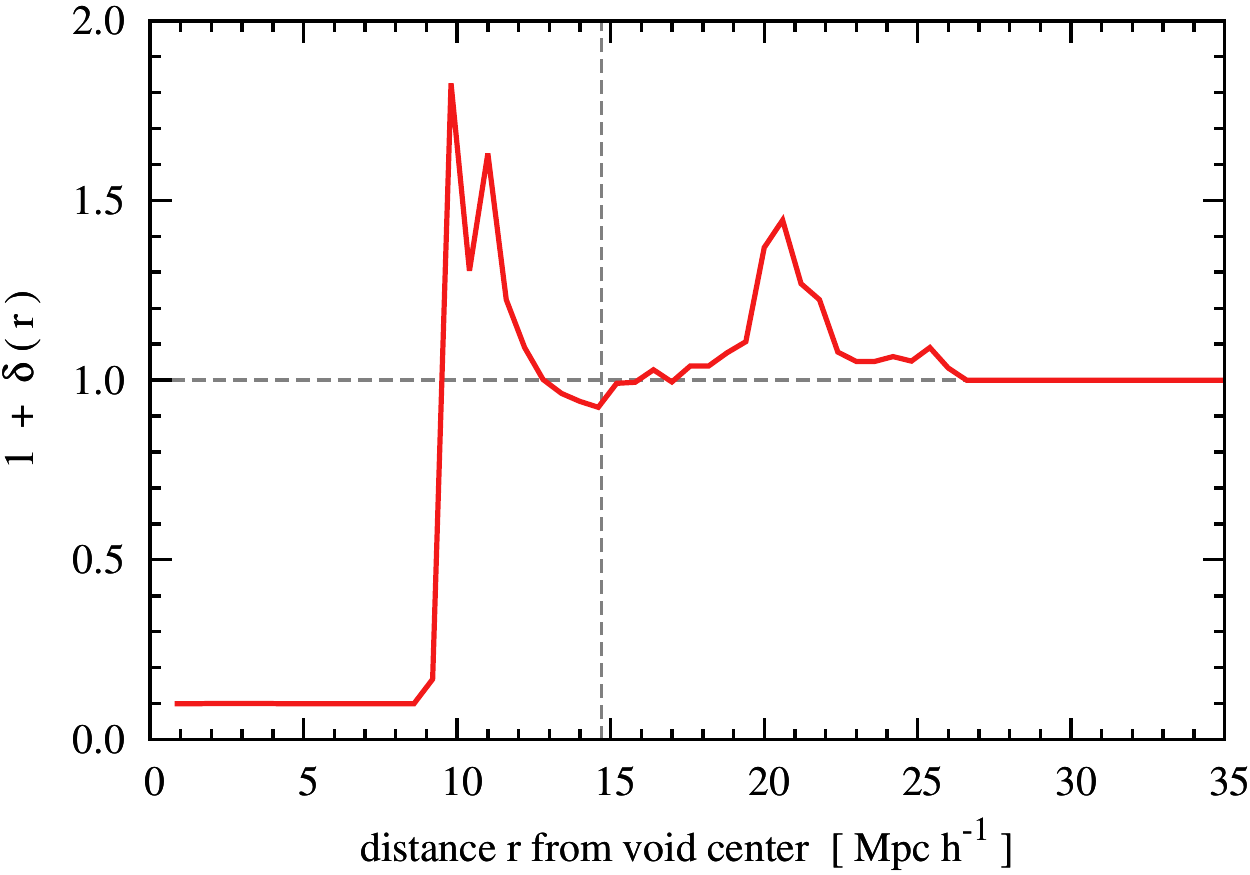} \\
        \includegraphics[width=.93\linewidth,angle=0]{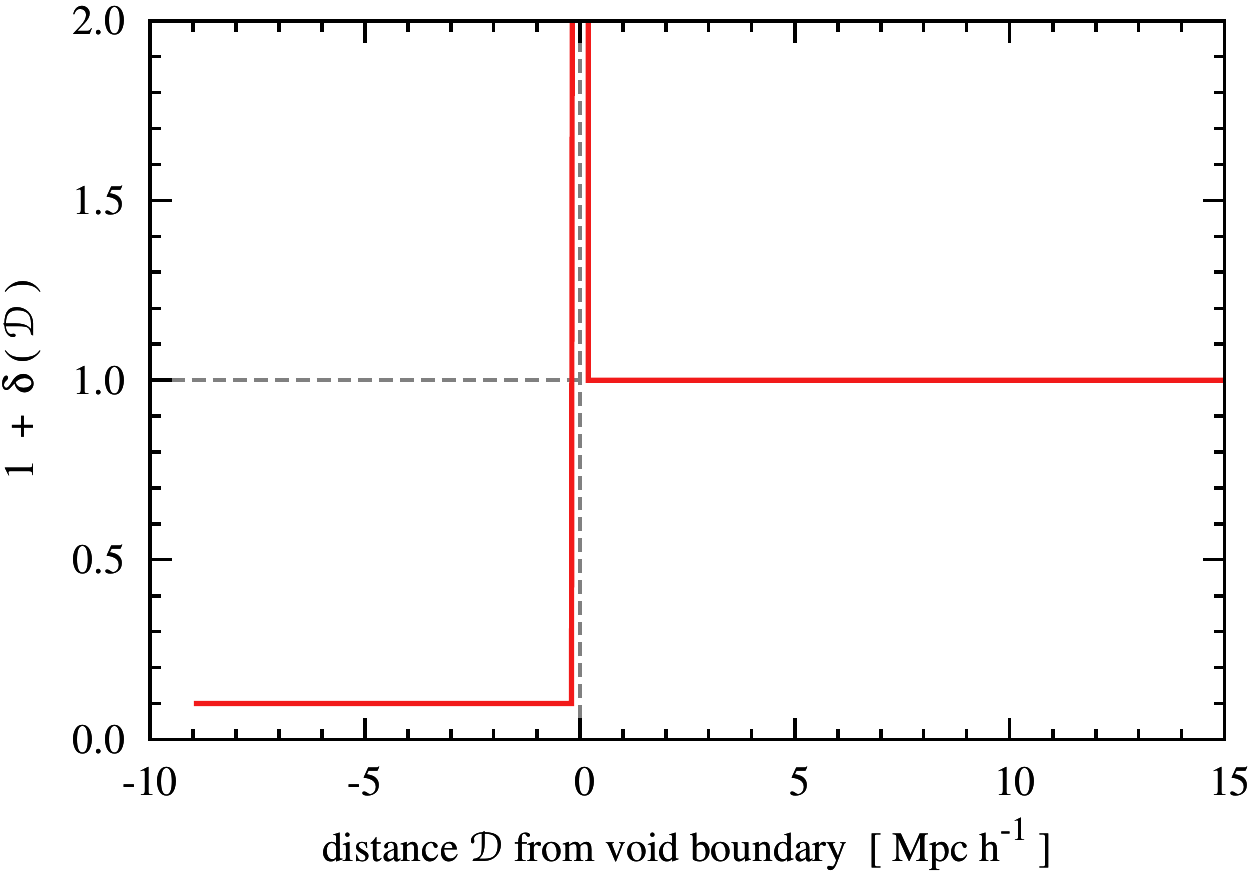}
    \end{array}$
     \caption{ Void density profile. The top panel shows the spherical profile of the simple void model illustrated in \reffig{fig:toy_example}. The vertical grey line marks the effective radius of the void, $\reff$, defined in \eq{eq:effective_radius}. The bottom panel shows the density profile of the void as a function of the distance, $\dist$, from the void boundary. For clarity, we define $\dist$ as having negative values inside the void and positive outside. The vertical grey line marks to the boundary of the void. }
     \label{fig:toy_example_profiles}
\end{figure}

Finding the \emph{spherically averaged} profile involves finding a void center, typically the volume-weighted barycentre,and growing concentric shells around it. This process is schematically illustrated in the centre panel of \reffig{fig:toy_example}, where, for clarity, we only show a few radial shells, but, in practice, we employ many more such shells. The spherical profile is given by the mean density of matter inside each shell. Applying this method to our model void provides the spherical density profile shown in the top panel of \reffig{fig:toy_example_profiles}. For small radial distances, which correspond to shells fully enclosed by the void, we recover the input density value, $1+\delta=0.1$. At larger radii, $r\ge9\Mpch$, the shells intersect the void boundary giving rise to a ``noisy" profile. Due to the irregular shape of the void, different radial shells have varying degrees of overlap with the void boundary, giving rise to ``noisy" features\footnote{In contrast to our simplified model, in real voids the mass is not distributed uniformly along the void boundary, resulting in even larger ``noisy" features.}. These persist for as long as the shells intersect the boundary, corresponding to $r\le26\Mpch$, while for even larger radii we recover the background density. This simple example illustrates that the spherical density profile is a complex convolution of the shape of the void and its actual density distribution.

To calculate the void profile with respect to the \emph{boundary} of the void we compute the boundary distance, $\dist{}$, that corresponds to the minimum distance from each point to the void boundary (see \eq{eq:void_distance} for a formal definition). The outcome is illustrated in the bottom panel of \reffig{fig:toy_example}, where each thin contour line corresponds to points that are at equal distance from the boundary of the model void. Now we can calculate the density profile as a function of $\dist{}$ by computing the mean density inside each shell of constant $\dist{}$ (in practice, we use many more shells than those shown in \reffig{fig:toy_example}). To distinguish between points inside and outside the void we adopt the convention that $\dist{}$ takes on negative values inside the void and positive outside, with $\dist{}=0$ at the void boundary. The resulting profile is plotted in the bottom panel of \reffig{fig:toy_example_profiles} and shows that we recover the actual input density distribution: $1+\delta=0.1$ inside the void, a large value of $1+\delta$ at the void boundary due to the mass evacuated from inside the void, and $1+\delta=1$ outside the void.

The new void profile has two major advantages. Firstly, it is independent of the shape of the void. For example, distorting the boundary of the void in \reffig{fig:toy_example}, while keeping the same density distribution inside and outside the void, would result in exactly the same density profile as a function of $\dist$. Secondly, on average, the mass displaced from inside the void is found at the void boundary, with the resulting density at the boundary being at least an order of magnitude higher than inside the void (\citealt{Sheth2004}, hereafter \SvdW{}, and \refsec{sec:density_profile}). Thus, while the spherical profile for radial shells that intersect the void boundary is dominated by the density at the boundary and not by the density inside the void, our proposed profile naturally differentiates between the boundary, the inside and the outside of the void.

%%%%%%%%%%%%%%%%%%%%%%%%%%%%%%%%%%%%%%%%%%%%%%%%%%%%%%%%%%%%%%%%%%%%%%%%%%%%%%%%%%%%
\section{Void identification}
\label{sec:data}

We make use of the high-resolution Millennium cosmological N-body
simulation \citep[\MI{};][]{Springel2005}. The \MI{} follows the evolution
of cold dark matter (DM) using $2160^3$ particles, each of mass, $m_p=8.6\times10^8\Msolar$, to resolve
structure formation in a periodic cube $500\Mpch$ on a side. 
%The spatial resolution (given by the Plummer-equivalent force softening) is, $\epsilon=1\kpch$, which was kept constant in comoving coordinates for the entire simulation. 
The \MI{} assumes the WMAP-1 cosmogony \citep{Spergel2003} with the following cosmological
parameters: $\Omega_m=0.23$, $\Omega_\Lambda=0.75$, $h=0.73$, $n_s=1$
and $\sigma_8=0.9$.

We identify voids using mock catalogues constructed from the semi-analytic galaxy formation model of \citet{Guo2011_SAM}. For this, we select only galaxies with stellar masses, $M_\star\ge3.8\times10^{10}\Msolar$, such that the number density is $n=3.2\times10^{-3}\MpchVolumeInverse{}$, similar to that of typical redshift surveys \citep[e.g.][]{Zehavi2011}. These galaxies are used as input to the Delaunay Tessellation Field Estimator \citep[DTFE;][]{Schaap2000,Weygaert2009,Cautun2011}, which uses a Delaunay triangulation with the galaxies at its vertices to extrapolate a volume filling density field. The resulting density field is used as input to the void identification method. We also apply the DTFE method to the distribution of DM particles to obtain continuous density and velocity fields, which are used for computing the density, velocity and weak lensing profile of voids. Both the galaxy density field and the DM density and velocity fields are stored on a $1280^3$ regular grid with a grid cell size of $0.39 \Mpch$.

The voids are determined using the Watershed Void Finder \citep[WVF;][]{Platen2007}, which identifies voids as the watershed basins of the large scale density field, similar to the ZOBOV void finder \citep{Neyrinck2008}. Compared to other methods, the watershed void finders have the advantage of not imposing any a priori constrains on the size, shape and mean underdensity of the voids they identify \citep{Colberg2008}. The WVF proceeds by first smoothing the galaxy density field with a $2~h^{-1} {\rm Mpc}$ Gaussian filter, whose size corresponds to the typical width of the filaments and walls forming the void boundaries \citep[e.g.][]{Cautun2013,Cautun2014a}. This smoothing is applied in order to dilute any substructures present on the void boundaries \citep[e.g. see][]{Cautun2014a}, which could potentially give rise to artificial voids. The smoothed density field is segmented into watershed basins using the watershed transform implemented using the steepest descent method \citep[e.g.][]{Bieniek2000}. This process is equivalent to following the path of a rain drop along a landscape: each volume element, in our case the voxel of a regular grid, is connected to the neighbour with the lowest density (i.e. steepest descent), with the same process repeated for each neighbour until a minimum of the density field is reached. Finally, a watershed basin is composed of all the voxels whose path ends at the same density minimum. 

\MCn{To overcome oversegmentation, the WVF joins the basins that share a boundary with a galaxy density, $\delta_{\rm g}\le-0.8$, since such low values typically separate subvoids embedded within larger voids. This threshold is motivated by the model of an expanding top-hat underdensity for which shell crossing takes place at $\delta=-0.8$ (\SvdW{}). This top-hat model has several shortcomings when compared to realistic voids, e.g. voids do not have initial top-hat profiles, their expansion is restricted by their environment and observations provide only the galaxy density, not the total matter density, which makes the extent to which $\delta_{\rm g}\le-0.8$ is a realistic threshold debatable.} \MCnn{This step leads to the merging of only $2\%$ of the watershed basins and hence it has no noticeable effect on the profiles of stacked voids.}

\begin{figure}
     \centering
     \includegraphics[width=\linewidth,angle=0]{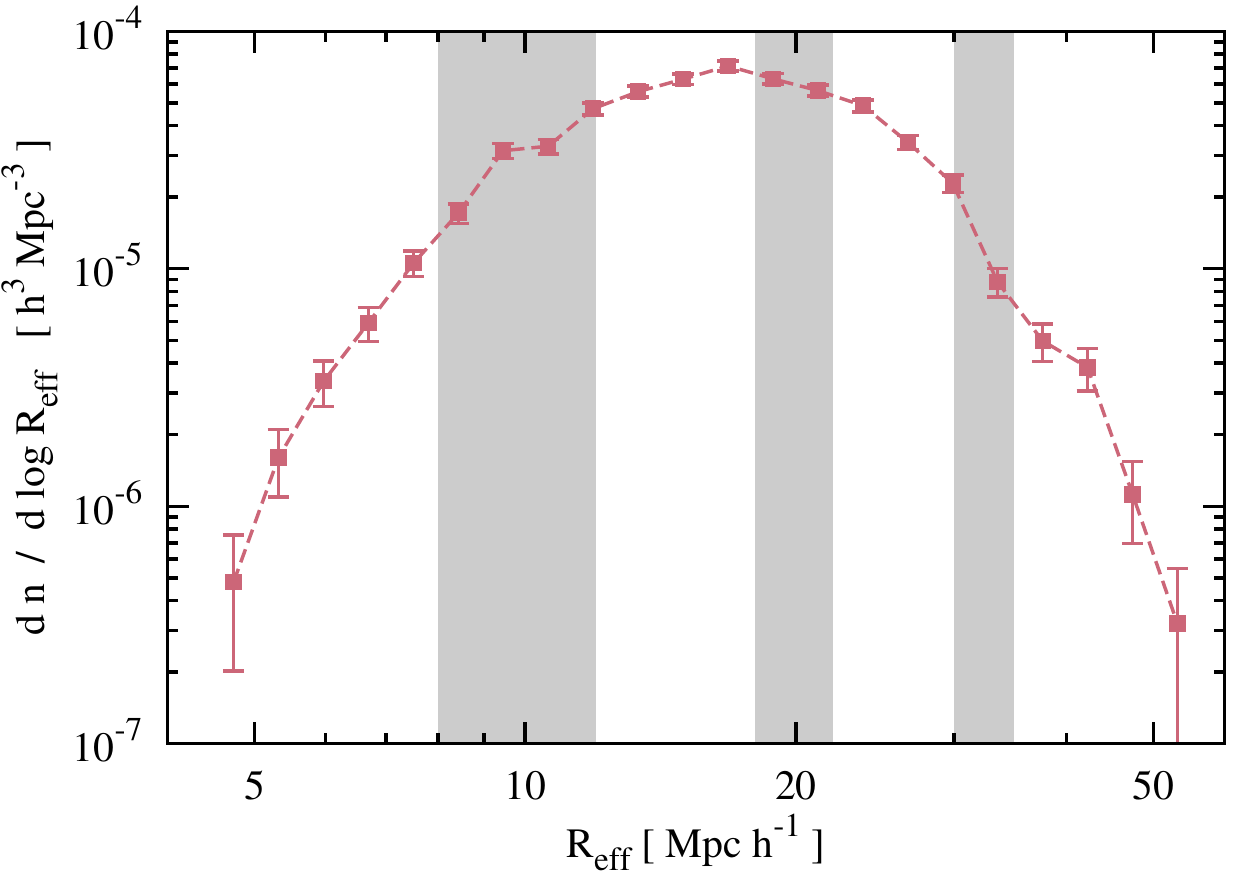}
     \caption{ The abundance of galaxy voids in the \MI{}. The shaded regions show the three ranges in effective void radius, $\reff{}$, for which we compute average density and velocity profiles. }
     \label{fig:void_abundance}
\end{figure}

The distribution of the WVF voids is shown in \reffig{fig:void_abundance} where the voids are characterised by their effective radius, $\reff{}$. This corresponds to the equivalent radius of a sphere with the same volume as the void, i.e.
\begin{equation}
    \reff{} = \left( \frac{3 V_{\rm void}}{4\pi} \right)^{\frac{1}{3}}
    \label{eq:effective_radius} \;,
\end{equation}
where $V_{\rm void}$ denotes the volume of the void. The figure shows that we identify a wide range of void sizes, from $5$ to $50\Mpch$, with the abundance peaking at $\reff\sim15\Mpch$. For the rest of this work, we will calculate stacked profiles for voids in three intervals in void size corresponding to $\reff=8-12$, $18-22$ and $30-35\Mpch$ (shown as dark shaded regions in \reffig{fig:void_abundance}), which contain 656, 643 and 100 voids, respectively.
These intervals were chosen to probe a variety of void sizes, while at the same time having enough voids to provide reliable statistics.

\MCn{The abundance of WVF voids is similar to that obtained using the ZOBOV void finder when applied to DM tracers with the same number density \citep[][Fig. 1]{Nadathur2015b}, but it is a factor of two higher than when applying ZOBOV to the galaxy distribution \citep[][Fig. 2]{Nadathur2015c}. The difference is likely due to the merging criteria employed by the two void finders (see \citealt{Nadathur2015c} who studied the dependence of the void abundance on merging criteria).} \MCnn{Regardless of these differences, the WVF voids have a similar minimum galaxy density to the ZOBOV voids \citep[see Fig. 1 in][]{Nadathur2015c}.}

\subsection{Spherical profiles}
\label{subsec:data_spherical_profiles}
The spherical profile of a void is computed as a function of the radial distance from the void centre, which we take as the volume-weighted barycentre. \MCn{While there are other potential choices of void centre  (e.g. see \citealt{Nadathur2015b}) that result in slightly different spherical profiles, these differences are small when compared to the difference between the spherical and the boundary profile methods.} The void centre is given by $\Vector{x}_{\rm vc}=\sum_{i}\Vector{x}_i/N$, where the sum is over all the $N$ voxels that are part of the void and $\Vector{x}_i$ gives the position of each such voxel. The density of the void at radial distance, $r$, is then given as
\begin{equation}
    \delta(r) = \frac{\sum_k w_k \delta_k}{\sum_k w_k}
    \label{eq:spherical_profile} \;,
\end{equation}
where the sum is over all the voxels found at a radial distance, $r\pm\tfrac{1}{2}\Delta r$, with $\Delta r$ the radial bin width, and $\delta_k$ is the density of each voxel. The weights, $w_k$, give the overlapping volume between the voxel and the radial bin. This is calculated by generating 64 points regularly distributed inside each voxel; $w_k$ is then the fraction of those points that are found inside the radial bin. This method of calculating void profiles is very similar to the VTFE method of \citet{Nadathur2015a}, except that we use a Delaunay instead of a Voronoi triangulation. Besides the density profile, we also compute the profile of the velocity component along the radial direction, $\vrad$. This is computed similarly to \eq{eq:spherical_profile}, but with the density replaced by the radial component of the velocity, $\vrad = \Vector{v} \cdot \Vector{r} / r$.

\subsection{Boundary profiles}
\label{subsec:data_shape_profiles}
To calculate the shape independent profile we need to identify the boundary or border of each void, which is the density ridge delimiting the watershed basin corresponding to that void. In practice, we compute the void boundary as follows. We loop over all the neighbouring grid cells of each voxel that is part of a void. If one of the neighbours is not part of the same void, then the face connecting the two voxels is identified as the boundary of the void. To speed up the computation, each such face is stored as only one point corresponding to the centre of the face. Finally, the border of the void is given by the union of all those points, i.e. by all the centres of the faces connecting voxels that are not part of the same void. This procedure can be easily expanded to identify the boundary of ZOBOV voids too. In this case, the voids are composed of Voronoi cells, not the cells of a regular grid as for the WVF. The boundary of ZOBOV voids is given by the union of the faces of the Voronoi cells that connect two Voronoi cells that are not part of the same void.

The next step is to compute the distance of each point from the void boundary, which in computer science is referred to as the distance transform. This technique has been previously used to find the galaxies that are the farthest inside voids \citep{Kreckel2011} and to use voids for improving photometric redshift estimation \citep{Aragon-Calvo2014}. The minimum distance from the void boundary to a point with Cartesian coordinate, $\Vector{x}$, is given by
\begin{equation}
    \dist = 
    \begin{cases}
        +\; \min_i\left[ \; \left| \Vector{x} - \Vector{y}_i \right| \; \right] & \text{for } \Vector{x} \text{ outside the void}\\
        -\; \min_i\left[ \; \left| \Vector{x} - \Vector{y}_i \right| \; \right] & \text{for } \Vector{x} \text{ inside the void}
    \end{cases}
    \label{eq:void_distance} \;,
\end{equation}
where $\{\Vector{y}_i\}$ denotes the set of points that give the void boundary and $|\,|$ denotes the magnitude of a vector. By convention, the boundary distance is negative for points inside the void and positive outside. One can further define a void boundary distance field, which at each point in space is a vector of magnitude, $|\dist{}|$, given by
\begin{equation}
    \distVec = \dist \frac{\Vector{x} - \Vector{y}_j}{\left| \Vector{x} - \Vector{y}_j \right|}
    \label{eq:void_distance_field} \;,
\end{equation}
where $j$ denotes the index of the point on the void boundary closest to $\Vector{x}$. The direction of $\distVec$ is perpendicular to the surfaces of constant $\dist$ (see bottom panel of \reffig{fig:toy_example}) and always points outwards. The void boundary distance field is computed separately for each void, using a kd-tree constructed from the set of points that gives the boundary of the void.

For each void, the boundary distance takes a minimum value, $\dist_{\rm min}$, that corresponds to the point inside the void farthest from the boundary. We always have $-\dist_{\rm min}\le\reff{}$ (note that $\dist_{\rm min}$ is negative), with equality only for spherical voids; on average, we find $\dist_{\rm min}/\reff{}=-(0.62\pm0.06)$ ($1\sigma$ standard deviation). For spherical voids, the boundary distance, $\dist$, is equivalent to the radial position, and, for this particular case, the spherical and boundary profiles are exactly the same.

The void density profile as a function of $\dist{}$ is computed similarly to \eq{eq:spherical_profile}:
\begin{equation}
    \delta(\dist{}) = \frac{\sum_k w_k \delta_k}{\sum_k w_k}
    \label{eq:distance_profile} \;,
\end{equation}
but now the sum is over voxels found at a distance $\dist{}\pm\tfrac{1}{2}\Delta\dist{}$ from the void boundary, with $\Delta\dist{}$ the width of the $\dist{}$ bin. The weight, $w_k$, is given by the fraction of the 64 uniformly distributed points inside each voxel that are within the required distance from the void boundary. As in \refsec{subsec:data_spherical_profiles}, we define the velocity component along the direction of $\distVec{}$ as $\vrad(\dist) = \Vector{v} \cdot \distVec / \dist$.

\subsection{Stacking}
\label{subsec:void_stacking}
The stacked profile is computed as an average over the individual profiles of voids in a narrow $\reff{}$ range. For spherical profiles, we average as a function of the rescaled radial distance, $r/\reff{}$. For \boundary{} profiles, we average over the individual voids at constant boundary distance, $\dist{}$. In this latter case, the minimum distance, $\dist_{\rm min}$, can differ even between voids of equal radii. As a result, at very low $\dist{}$ values only a subset of the voids contribute to the stacked \boundary{} profile. This effect, which our calculation accounts for, becomes important only for the lowest $\dist{}$ bins and can be easily spotted due to the large error bars associated with these points. The uncertainty for the two stacking procedures is given as the $1\sigma$ interval in the distribution of the mean value obtained using 1000 bootstrap samples.

%%%%%%%%%%%%%%%%%%%%%%%%%%%%%%%%%%%%%%%%%%%%%%%%%%%%%%%%%%%%%%%%%%%%%%%%%%%%%%%%%%%%%%%%
\section{Void density profile}
\label{sec:density_profile}
We now apply the \boundary{} profile analysis to galaxy voids found in the \MI{}. To demonstrate the power of this new method, we compare with the outcome of the conventional approach based on spherically averaged profiles. 

\subsection{Individual voids}
\begin{figure}
     \centering
    $\begin{array}{c}
        \includegraphics[width=0.95\linewidth,angle=0]{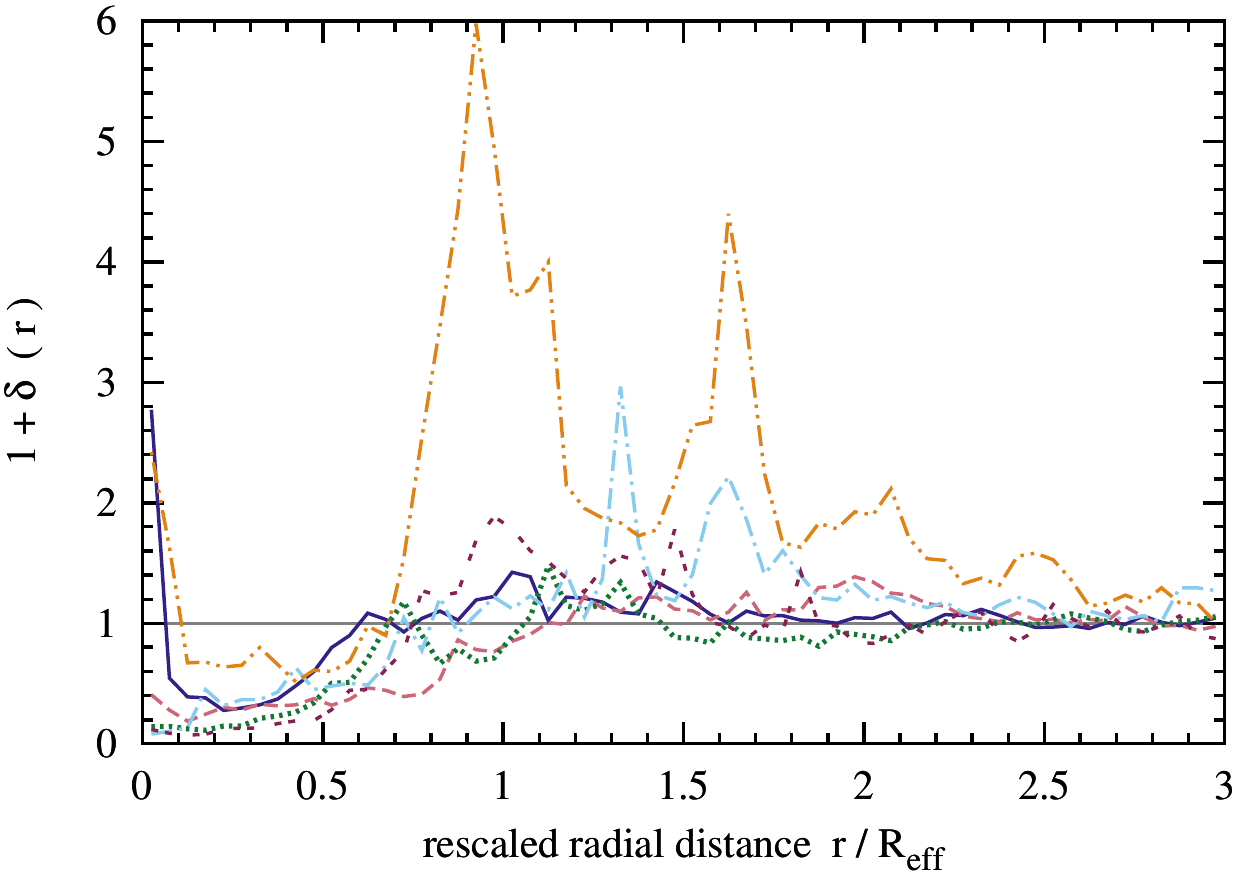} \\
        \includegraphics[width=0.95\linewidth,angle=0]{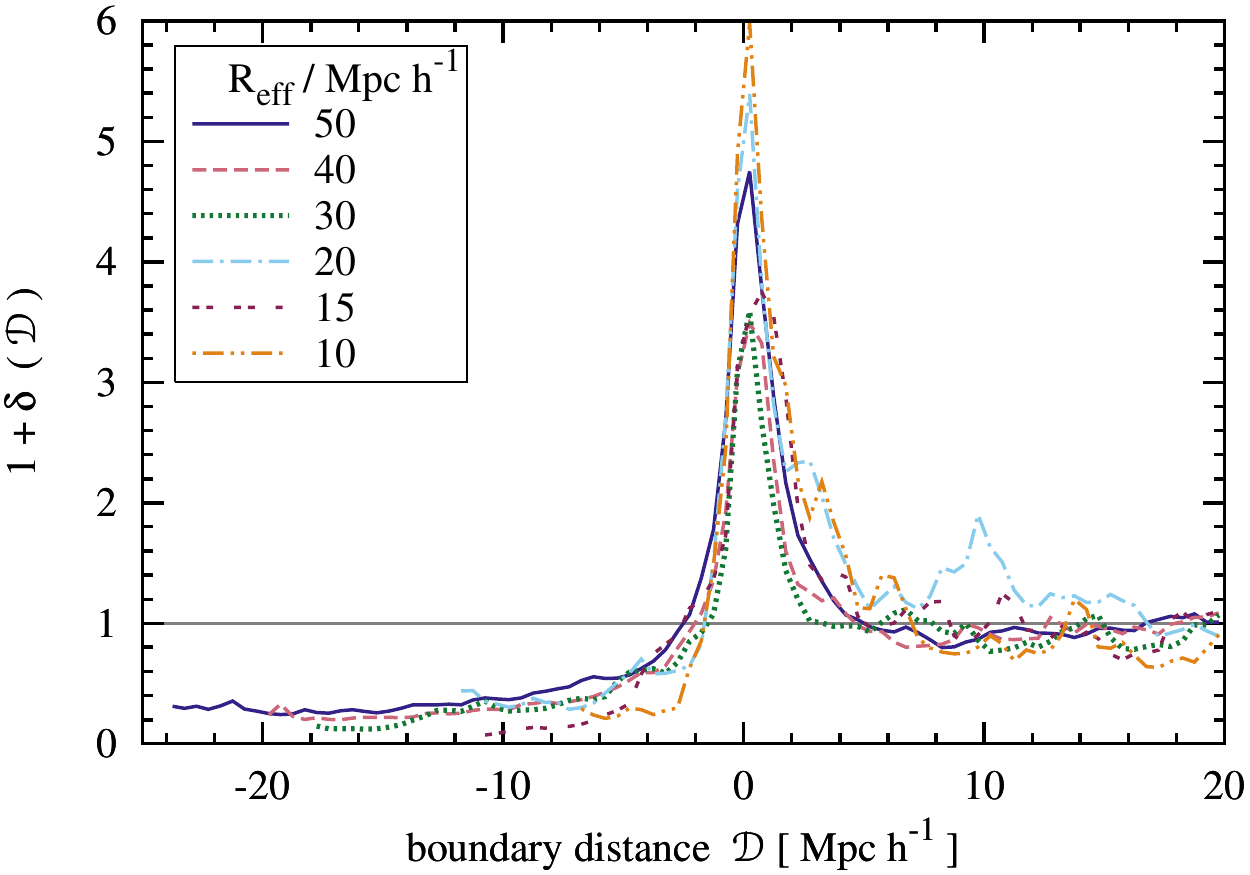}
    \end{array}$
     \caption{ The density, $1+\delta$, profile of six randomly selected voids that span a range of effective void sizes, $\reff{}$. The top panel shows the spherical profile as a function of the rescaled radial position, $r/\reff{}$. The bottom panel shows the \boundary{} profile as a function of the distance, $\dist{}$, from the boundary of the void. }
     \label{fig:density_profiles_random_voids}
\end{figure}

\begin{figure}
     \centering
    $\begin{array}{c}
        \includegraphics[width=0.95\linewidth,angle=0]{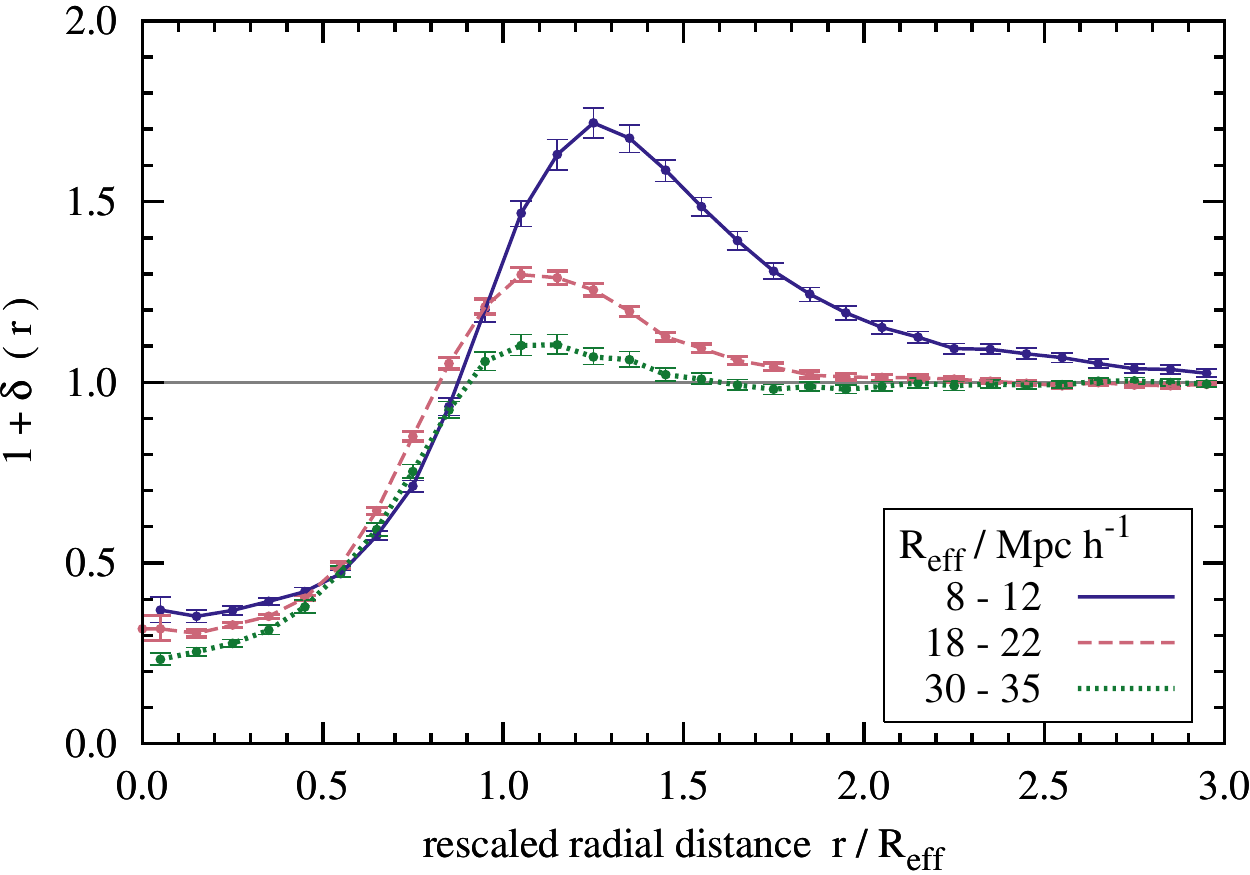} \\
        \includegraphics[width=0.95\linewidth,angle=0]{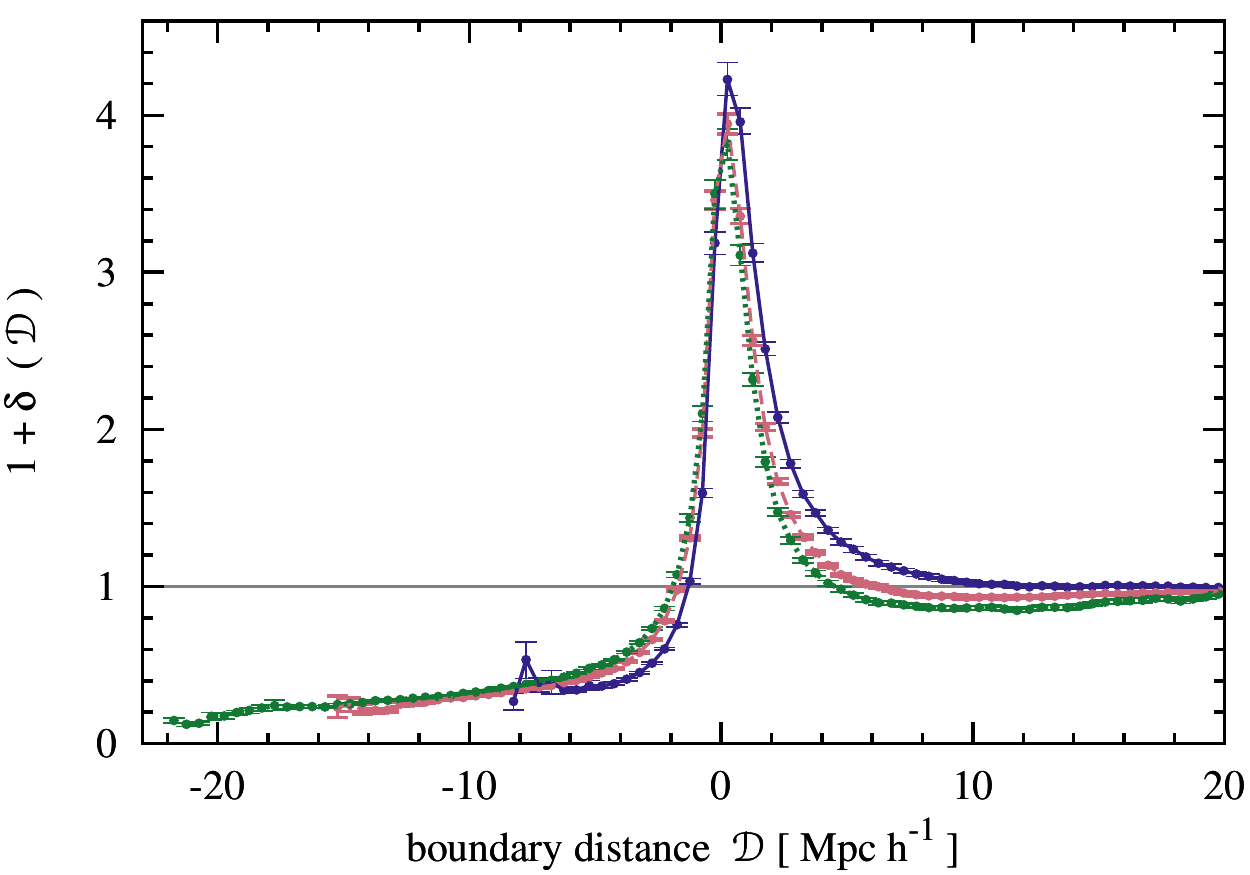}
    \end{array}$
     \caption{ The stacked density profile for voids in three ranges in effective radius, $\reff{}$. %with sizes $\reff{}=$ $8-12$, $18-22$ and $30-35\Mpch$. 
     The two panels show the spherical (top) and the \boundary{} (bottom) profiles. }
     \label{fig:density_profiles_stacked}
\end{figure}

In \reffig{fig:density_profiles_random_voids} we show the density profile of six random voids selected to span a wide range of sizes. These profiles and the subsequent ones are computed using the full DM particle distribution and hence give the overdensity of matter. The \emph{spherical} profile is shown as a function of the rescaled radial distance, $r/\reff{}$, which is used for determining self-similar and universal void profiles \citep{Hamaus2014b,Nadathur2015a}. While there is a large variation between the different voids and between neighbouring radial bins, on average individual voids are underdense for $r/\reff{}\lsim0.5$ and show no consistent features at larger distances. Two of the voids have $\delta>0$ at their centres that can be explained either by the presence of a substructure inside the void \citep{Beygu2013,Rieder2013} or by the void centre being close to the boundary \citep{Nadathur2015b}.

Compared to the spherical profile, the \emph{\boundary{}} profile is very different and has more features, indicative of the fact that, since voids have highly complex shapes, taking a spherical average erases or damps many features. In addition, the \boundary{} profile shows a better qualitative agreement between the various voids: underdense for $\dist\le-3\Mpch$; a sharp density peak at the void boundary, $\dist=0$; and close to mean background density for $\dist\ge5\Mpch$. The density peak at the void boundary is expected since voids identified using watershed-based methods are delimited by a density ridge. The height of this ridge is largely given by the mass contained in the most massive haloes, which explains the variation in height between different voids. Massive haloes can also be found outside the void, resulting in sporadic peaks in the density profile, but only very rarely inside the void - no such example is present in \reffig{fig:density_profiles_random_voids}. The width of the density peak at the boundary is given by the typical size of the massive haloes as well as that of the filaments and walls that delimit the void \citep[e.g.][]{Cautun2013,Cautun2014a}.

\subsection{Stacked profiles}
In \reffig{fig:density_profiles_stacked} we present the mean density profiles of voids in three $\reff{}$ bins chosen to probe a variety of void sizes (see \reffig{fig:void_abundance}). The \emph{spherical} stacked profiles are underdense in the inner parts, with $\delta$ slowly rising to a maximum at $r\simeq\reff$, followed by a gradual transition towards the average background density \citep{Hamaus2014b,Nadathur2015a}. 
% The majority of small voids correspond to largely collapsing voids embedded in overdense regions, while the largest objects
%The smaller voids have a large density peak and correspond to largely collapsing voids embedded in overdense regions. \MCn{The height of the density peak decreases with $\reff{}$, with the largest objects corresponding to mostly expanding voids \citep{Paz2013,Hamaus2014c}.}

The \emph{\boundary{}} profile paints a different picture of the structure of voids. In the inner most parts, $\dist\lsim-4\Mpch$, the density is very low, $-0.9\le\delta\le-0.5$, and nearly constant, with only a very small increase in $\delta$ with $\dist$. This is followed by a very steep rise of a density ridge at the boundary, which decreases nearly as fast at $\dist\ge0$. At even further distances, the density gradually reaches the background value. 

\begin{figure}
     \centering
     \includegraphics[width=0.7\linewidth,angle=0]{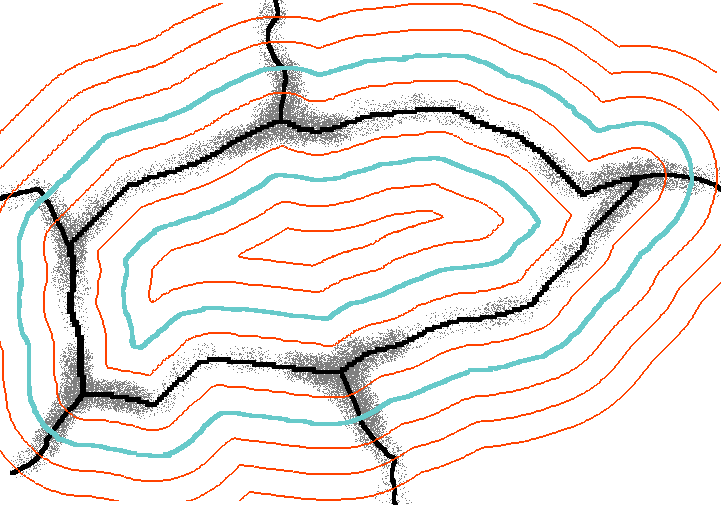} 
     \caption{ A simple model to understand the \boundary{} profile. The thick black curves show the boundary of the central void and that of its neighbours, which are coloured according to their density, with dark and light grey showing high and low density. The highest density regions correspond to the intersection points of two or more void boundaries, with the density decreasing farther away. The thin curves shows contours of constant distance, $\dist$, from the boundary of the central void, with two of those contours, $\dist=-5$ and $5\Mpch$, highlighted in cyan. The outer contours intersect the boundary of neighbouring voids and hence correspond to a higher mean density than the inner contours. }
     %Explaining the asymmetry between the inner and outer parts of the \boundary{} profile. The thick black curves show the boundary of the central void and that of its neighbours. The thin curves shows contours of constant distance, $\dist$, from the boundary of the central void, with two of those contours, $\dist=-5$ and $5\Mpch$, highlighted in cyan. The outer contours intersect the boundary of neighbouring voids and hence correspond to a higher mean density than the inner contours. }
     \label{fig:void_neighbours}
\end{figure}

The \boundary{} density profile can be understood within the multiscale picture of the cosmic web. Void interiors are not fully empty, but instead are criss-crossed by tenuous filaments and walls that become more densely packed as one approaches the massive structures that delimit the voids \citep{Cautun2014a}. Thus, the mean density is expected to increase close to the void boundary, in accord with the results shown in the bottom panel of \reffig{fig:density_profiles_stacked}. Close to the void boundary, the behaviour is dominated by the prominent filaments and sheets that delimit the void and that are substantially denser than the tenuous structures found inside the void \citep{Cautun2014a}. The picture outside the void is complicated by the presence of neighbouring voids and their own dense ridges, as illustrated in \reffig{fig:void_neighbours}. The density profile is not symmetric around $\dist{}=0$ since neighbouring voids can have different sizes, and hence different ridge thicknesses. In addition, the outer contours intersect the boundary of neighbouring voids. Due to clustering, the density varies along the void ridge, with higher density typically associated with the intersection points of two or more void boundaries. The density profile is sensitive to this clustering, which would explain why the slope, $\left|\tfrac{d\delta}{d\dist}\right|$, is shallower outside the void than inside the void.

Compared to the spherical profile, the average \boundary{} profile shows smaller differences between voids of different sizes and is close to a self-similar profile. Before discussing these differences, we proceed by fitting the \boundary{} profile with the empirical function:
\begin{equation}
    \rho = 
    \begin{cases}
        \rho_{\rm in}   \left( 1 + \left(\frac{\rho_{\rm max}}{\rho_{\rm in}}  -1\right) e^{-\frac{|\dist|}{t_{\rm in}}}   \right) (1{-}\alpha|\dist|)
                 & \text{for } \dist<0\\
        \rho_{\rm out}  \left( 1 + \left(\frac{\rho_{\rm max}}{\rho_{\rm out}}-1\right) e^{-\frac{|\dist|}{t_{\rm out}}} \right)
                & \text{for } \dist\ge0
    \end{cases}
    \label{eq:fit_function_density} \;,
\end{equation}
where \MCn{$\rho = {\bar\rho} (1+\delta)$ is the matter density and ${\bar\rho}$ is the mean background density}. The fit is a continuous function composed of two parts that describe the inner, $\dist<0$, and outer, $\dist\ge0$, mean density profiles, with $\rho(\dist{}=0)=\rho_{\rm max}$.

\begin{figure}
     \centering
     \includegraphics[width=\linewidth,angle=0]{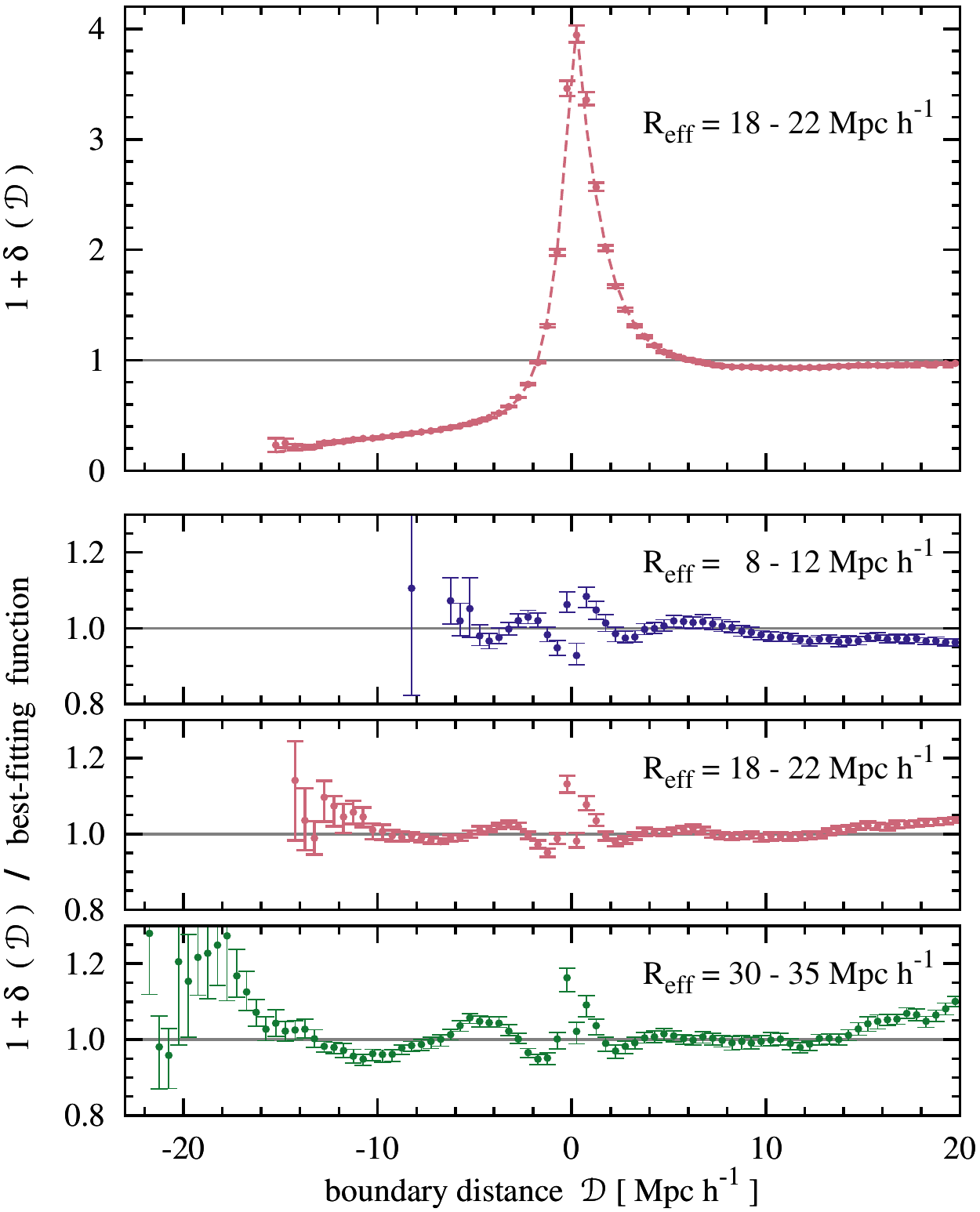} 
     \caption{ The best-fitting function (Eq. \ref{eq:fit_function_density}) to the \boundary{} density profile of voids. The top panel shows the mean density for voids with $\reff=18-22\Mpch$ (symbols with error bars) and the best-fit function (dashed line). The remaining panels show the ratio between the data and the best-fitting function for voids of different sizes. The fit was done using only data points with $\dist\le10\Mpch$.}
     \label{fig:density_profile_fit}
\end{figure}

The very interiors of the void are characterised by the density parameter, $\rho_{\rm in}$, and by the slope $\alpha$, the latter accounting for the fact that the density increases with $\dist{}$. The density ridge at $\dist\simeq0$ is well described by an exponential function that takes a maximum value, $\rho_{\rm max}$, at $\dist=0$. This ridge is not symmetric with respect to $\dist=0$ and so we have two parameters in the exponential, $t_{\rm in}$ and $t_{\rm out}$, that give the thickness of the inner and outer void boundary, respectively. Just outside the void boundary, the density has yet to converge to the background value, so there is an additional parameter, $\rho_{\rm out}$, to account for this effect. The $\dist{}\ge0$ part of the fitting function should include an additional component to account for the transition towards the background density at large $\dist$, but, for simplicity, we omit such a component. Our function is characterised by six parameters which is similar to other empirical fits to spherical void profiles: \citet{Hamaus2014b} proposed a four parameter fit that was latter extended by \citet{Barreira2015} to a five parameter fit to give a better description of voids identified in the DM density field. Compared to the \boundary{} profile, the spherical one smooths over many density features, so it is not surprising that the former requires more parameters.

\begin{figure}
     \centering
     \includegraphics[width=1.\linewidth,angle=0]{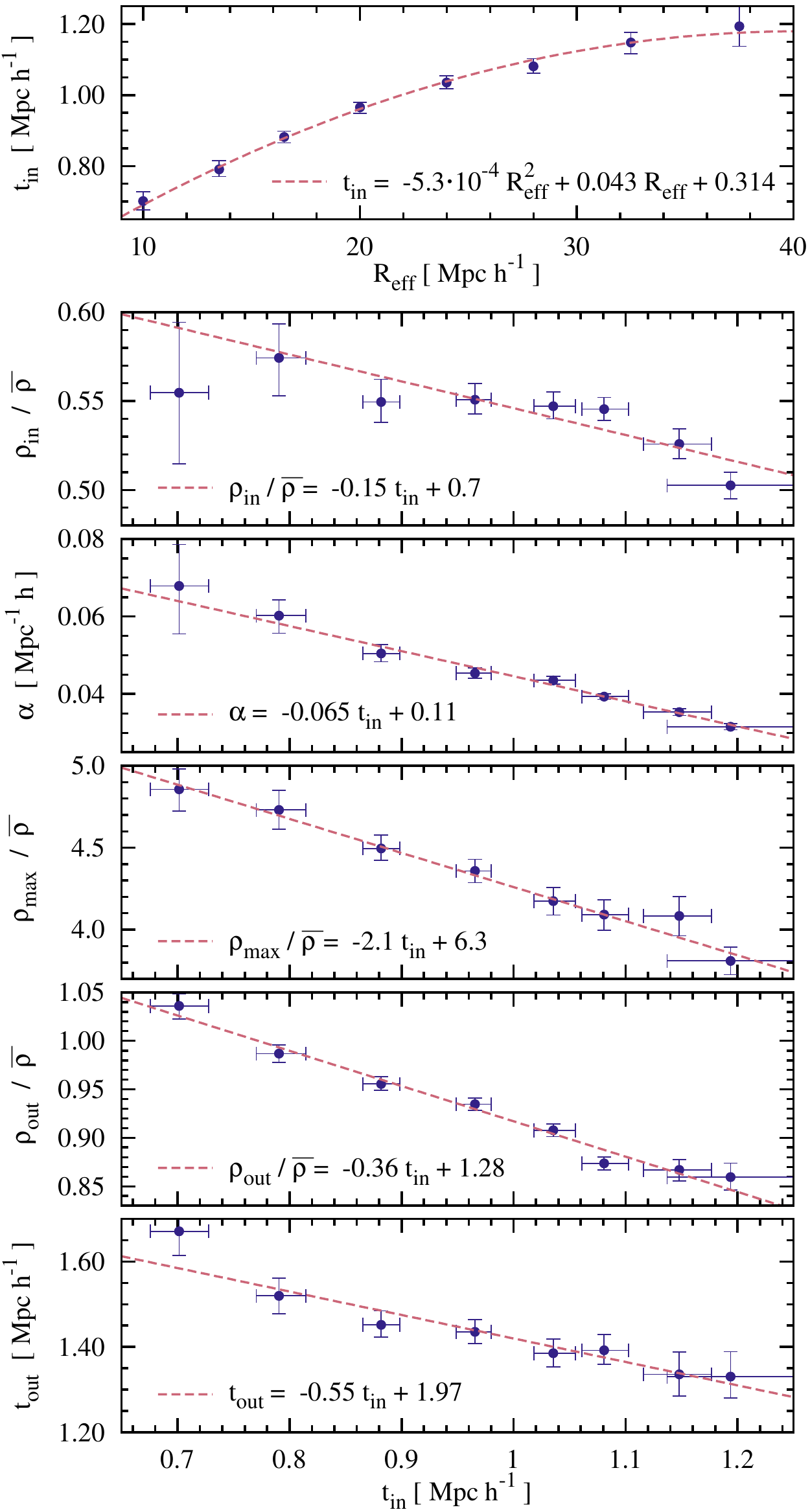} 
     \caption{ The best-fitting parameters of \eq{eq:fit_function_density} obtained from stacked void density profiles. The top panel show the thickness of the inner void boundary, $\tin$, as a function of void radius. The remaining panels show the dependence of the other fit parameters: $\rho_{\rm in}$, $\alpha$, $\rho_{\rm max}$, $\rho_{\rm out}$ and $\tout$ as a function of $\tin{}$. The error bars give the $1\sigma$ uncertainty. The dashed lines show that the best-fitting parameters follow simple relations with $\reff{}$ (top panel) and $\tin{}$ (remaining panels). }
     \label{fig:best-fitting_parameters}
\end{figure}

The upper panel of \reffig{fig:density_profile_fit} shows that the empirical function of \eq{eq:fit_function_density} describes, to very good approximation, the mean density profile. To better assess the fit quality, the lower panels of \reffig{fig:density_profile_fit} show the ratio between the measured profile and the best-fitting value for three void samples. The fit matches the data well, except for a few points around $\dist\simeq0$, which show a ${\sim}10\%$ difference, and for the $D\le-15\Mpch$ region of the largest voids, which shows a systematic deviation from the best-fit.

\reffig{fig:best-fitting_parameters} shows the best-fitting parameters and their $1\sigma$ errors for voids of different size. These were computed using the Markov chain Monte Carlo method implemented in the emcee package \citep{Foreman-Mackey2013}. The figure shows $\tin{}$ as a function of $\reff{}$ and the remaining parameters as a function of $\tin{}$. The best-fitting parameters follow linear relations with $\tin{}$, which in turn can be parametrized as a quadratic function of $\reff{}$. This suggests that the parametrization of \eq{eq:fit_function_density} is overdetermined and that the number of free parameters is too large \citep[similar relations between the fit parameters have been reported by][]{Hamaus2014b}. \eq{eq:fit_function_density} can be rewritten by expressing $\rho_{\rm in}$, $\alpha$, $\rho_{\rm max}$, $\rho_{\rm out}$ and $\tout$ as a linear function of $\tin{}$ (two parameters in each case) and, in turn, by expressing $\tin{}$ as a quadratic function of $\reff{}$ (three parameters). This results in a 13 parameter function that fits in one step voids of various sizes. We repeated the fit using these parametrizations and obtained similarly good fits.

According to \reffig{fig:best-fitting_parameters}, void interiors are characterised by a nearly constant density, $\rho_{\rm in}$, but by different values of the density slope, $\alpha$, with larger voids having more slowly varying density profiles. The height of the density ridge, $\rho_{\rm max}$, is largest for small voids since these are typically embedded in overdense regions. This is illustrated also by the $\rho_{\rm out}/\bar\rho$ density parameter that is larger than $1$ for the smallest voids and that decreases with void size. The density ridge is asymmetric and is thinner inside the void, i.e. $t_{\rm in}<t_{\rm out}$ (see the discussion of \reffig{fig:void_neighbours}). 

We also find that the smallest voids have lower $t_{\rm in}$ values and larger $t_{\rm out}$ values than the largest voids. The increase of $t_{\rm in}$ and decrease of $t_{\rm out}$ with void size can be a manifestation of the age characterising voids of different size. Just as low mass haloes, small voids are dynamically old, so the density ridge has been squeezed for a longer time. Larger voids, which originate from larger scale density fluctuations, have not had enough time to pile up mass at the ridge to the same extent as the small ones. 
%Secondly, small voids are predominantly contracting, which means that, on average, the outer shells are moving inward while the inner shells are moving outwards, leading to sharper ridges. In contrast, large voids are expanding, with their outer shells systematically moving outwards (see \refsec{sec:velocity_profile} where we study the velocity profile). Such an effect would also lead to a thinner inner ridge for small voids and a thicker one for large voids.

\subsection{The self-similarity of stacked profiles}
\label{subsec:self-similarity}
\begin{figure}
     \centering
     \includegraphics[width=\linewidth,angle=0]{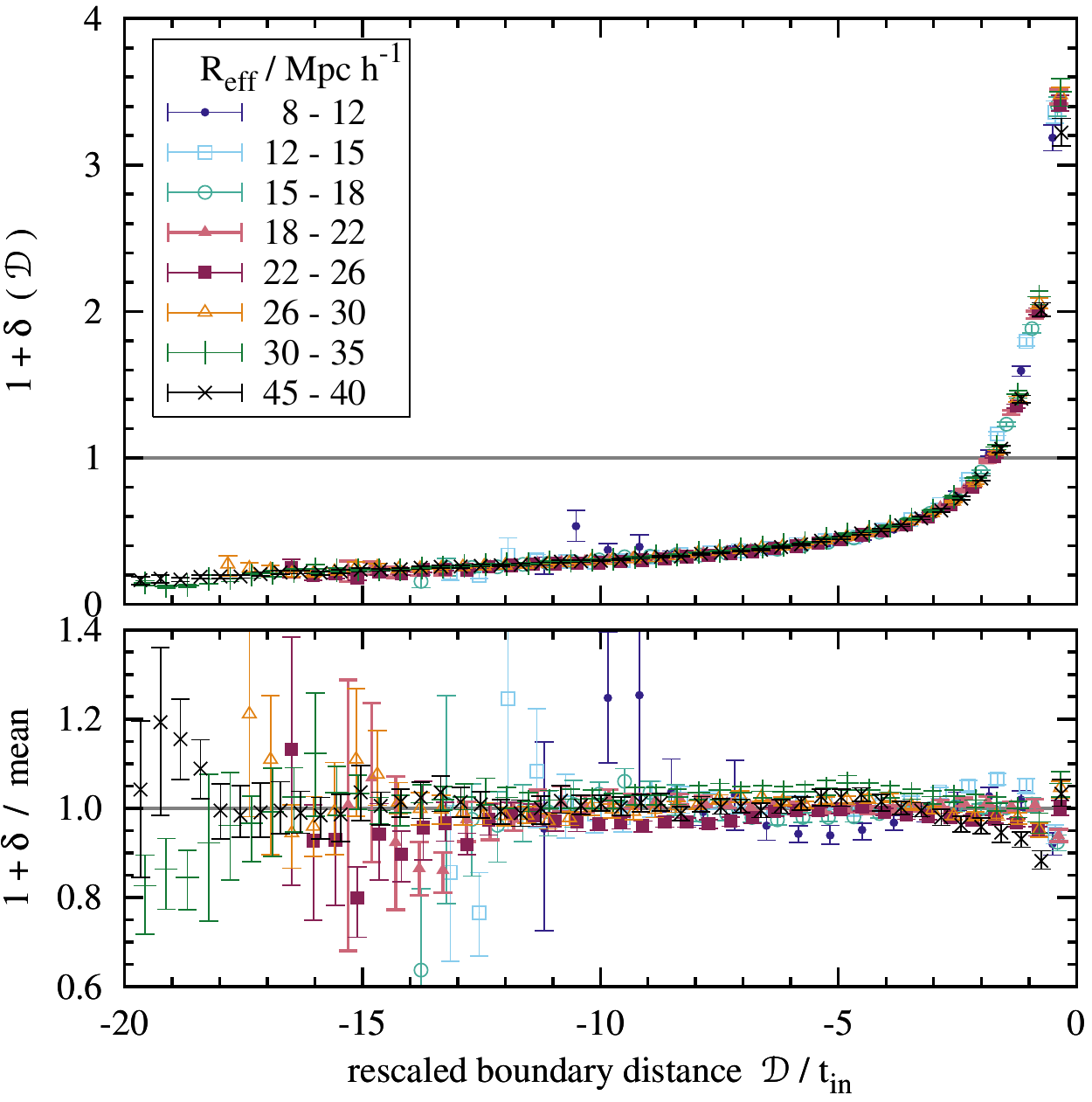}
     \caption{ The self-similarity of voids. Top panel: the density profile, $1+\delta$, as a function of the rescaled void boundary distance, $\dist/\tin$, where $\tin{}$ is the thickness of the inner void boundary as determined by fitting \eq{eq:fit_function_density} to the density profile. The symbols correspond to voids of various effective radii, $\reff{}$. All voids have a self-similar density profile independent of $\reff{}$. Bottom panel: the ratio between the profiles and a weighted mean of the values in the various $\reff{}$ bins showing that there is less than $5\%$ scatter among voids of various sizes. }
     \label{fig:density_profiles_scaling}
\end{figure}

The \boundary{} density profile of voids of different size is very similar, but not exactly the same (see bottom panel of \reffig{fig:density_profiles_stacked}). Those differences are minimized, or even disappear entirely, when rescaling the inner profile by the thickness, $t_{\rm in}$, of the inner void boundary. The rescaled profiles are given in \reffig{fig:density_profiles_scaling} which clearly shows that all voids, independently of their size, have a self-similar profile. To better highlight this, in the bottom panel of the figure we take the ratio with respect to a weighted mean density. This weighted mean was obtained by averaging, at fixed $\dist/t_{\rm in}$ values, over voids of different sizes, with the contribution of each sample weighted by the inverse of its associated error. Small systematic differences with void size are seen only for $\dist\simeq0$, which probably arise because small void are embedded in overdense regions while large voids are found in predominantly underdense regions. For the rest, all the density profiles lie on the same curve with less than ${\sim}5\%$ scatter.

The self-similar nature of \boundary{} profiles suggest that voids of different sizes have, on average, the same dynamical characteristics. In contrast, the same self-similarity is not seen for spherical profiles (see top panel of \reffig{fig:density_profiles_stacked}). This could be due to the limitations of spherical profiles, among which, most importantly, is the mixing and inability to separate between the inside, boundary and outside of voids, as we exemplified in \refsec{sec:whats_about}. This fits in with the results shown in the bottom panel of \reffig{fig:density_profiles_stacked} where the differences between voids of various sizes are most pronounced in the boundary and outside regions of the voids. 

Self-similar profiles are obtained only after rescaling by the thickness of the inner void ridge, $t_{\rm in}$. This suggests that the void interior knows about the boundary or vice-versa, and that the two evolve together. The former possibility seems ruled out by the simple picture of an expanding spherical underdensity in which the evolution of a shell of matter of radius, $r$, depends only on the mass contained within $r$ \citep[][\SvdW{}, but see \citealt{Ruiz2015}]{Fillmore1984}.

Spherical void profiles have also been claimed to be self-similar \citep[e.g.]{Ricciardelli2014,Nadathur2015a}, but there are contradictory results in the literature \citep[e.g.][this work]{Hamaus2014b,Sutter2014b,Nadathur2015b}. The self-similarity of spherical profiles seems to be dependent on several factors: the void finder, the population of tracers used to identify the voids and the tracers used to measure the void profile. This could be the case for \boundary{} profiles too, though it is reassuring that self-similarity of \boundary{} profiles has been found for both voids identified using galaxies (this work) and for voids identified in the DM density field \citep{Cautun2015e}.

\subsection{Comparison to analytical predictions}
\begin{figure}
     \centering
     \includegraphics[width=\linewidth,angle=0]{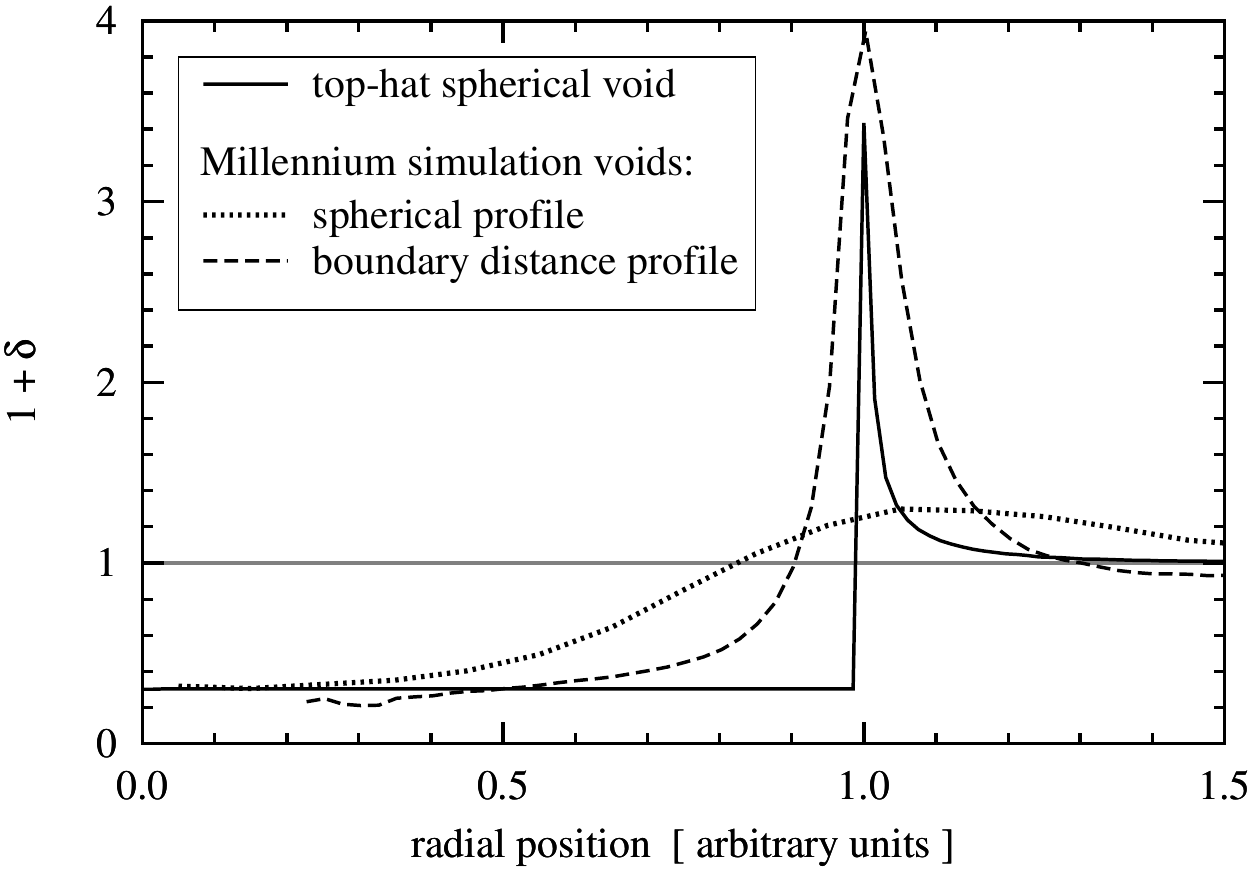}
     \caption{ Comparison of analytical and measured density profiles of voids. \MCn{The solid line corresponds to an uncompensated top-hat spherical underdensity that gives rise to a void with mean density, $1+\delta=0.3$.} The dotted and dashed curves give the spherical and \boundary{} distance profiles of \MI{} voids with $\reff=18-22\Mpch$. The top-hat void shows a good qualitative agreement with the boundary distance profile of \MI{} voids. }
     \label{fig:density_profiles_top-hat}
\end{figure}

It is illustrative to compare with analytical predictions of void profiles, among which the isolated spherical underdensity model is the most popular (\citealt{Fillmore1984}, \SvdW{}; see \citealt{vandeWeygaert2011} for a more elaborate void evolution model that includes ellipsoidal collapse and that accounts for the effect of the external tidal field). For this purpose, we select a top-hat spherical underdensity that \MCn{gives rise to a void of radius, $\reff{}=20\Mpch$, and density, $1+\delta=-0.3$, similar to the mean density of stacked \MI{} voids with sizes, $\reff=18-22\Mpch$.} While realistic voids do not have initial top-hat profiles, such a simple model captures most of the features of initial underdensities representative of cosmological environments (see Fig. 3 of \SvdW{}). \reffig{fig:density_profiles_top-hat} shows the density profile of the resulting void as a function of the rescaled radial distance, $r/\reff{}$. The figure also shows the spherical and \boundary{} profile of \MI{} voids with sizes, $\reff=18-22\Mpch$. To plot all three profiles on the same x-axis, we give the \boundary{} profile in terms of the rescaled coordinate, $(\dist+\reff)/\reff$, with $\reff{}=20\Mpch$. 

The top-hat profile shows large differences with respect to the spherical profile of \MI{} voids, but is in approximate agreement with the \boundary{} profile of the same voids. In particular, the \boundary{} profile matches the main prediction of the analytical model, the formation of a density ridge at the edge of the void. Thus, this simple model offers a qualitative description of the density profiles of voids, but only after accounting for the fact that real voids are non-spherical.

\MCn{Note, however, that there are significant differences between the top-hat model and the \boundary{} density profile of realistic voids, which are driven by many factors. Our goal is not to test the accuracy of the analytical model, but rather to show that such a model performs better than one would naively expect from a comparison to spherical profiles. For example, the edge of \MI{} voids contains more mass than the analytical prediction since the boundaries of realistic voids accrete matter also from outside the void (note the $1+\delta<1$ values of the \boundary{} profile for rescaled radial positions larger than $1.3$). Secondly, replacing the uniform top-hat underdensity with more realistic initial density profiles results in a more gradual increase of the density ridge (\SvdW{}), which is closer to the profile of \MI{} voids.}
 
\reffig{fig:density_profiles_top-hat} is a first step towards testing one of the central assumptions of the \SvdW{} void abundance model, which is that void formation is well described by the evolution of \MCnnn{an isolated spherical underdensity}. We have shown that we can find \emph{a} top-hat model that qualitatively matches the \emph{mean} density of stacked voids. It remains to be seen if the parameters of this top-hat model are also the ones required to match the initial conditions of realistic voids. Furthermore, for the model to be realistic, the match should work not only for stacked samples, but also for individual voids.

%%%%%%%%%%%%%%%%%%%%%%%%%%%%%%%%%%%%%%%%%%%%%%%%%%%%%%%%%%%%%%%%%%%%%%%%%%%%%%%%%%%%%%%%
\section{Void velocity profile}
\label{sec:velocity_profile}

The velocity field of voids is another property that can be better understood by analysing \boundary{} profiles. As for the density profile, we proceed by comparing the spherical and \boundary{} velocity profiles. We focus on the peculiar velocity component, $\vrad$, that gives the rate at which matter is evacuated in comoving coordinates through a surface of $r=\textrm{const}$ and $\dist=\textrm{const}$ for the spherical and \boundary{} profiles, respectively. Positive $\vrad$ values correspond to a net outflow of matter through the surface while negative values correspond to an inflow.

\begin{figure}
     \centering
    $\begin{array}{c}
        \includegraphics[width=\linewidth,angle=0]{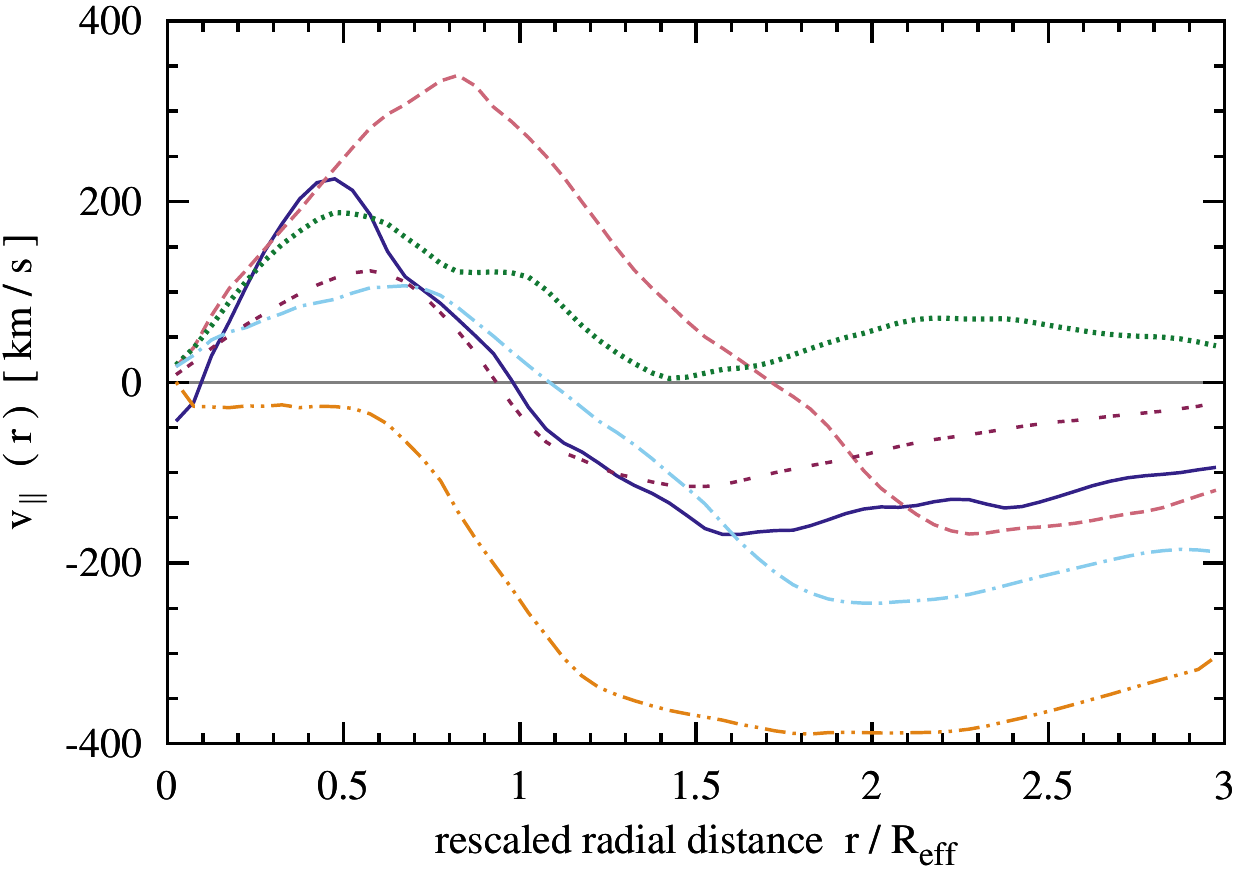} \\
        \includegraphics[width=\linewidth,angle=0]{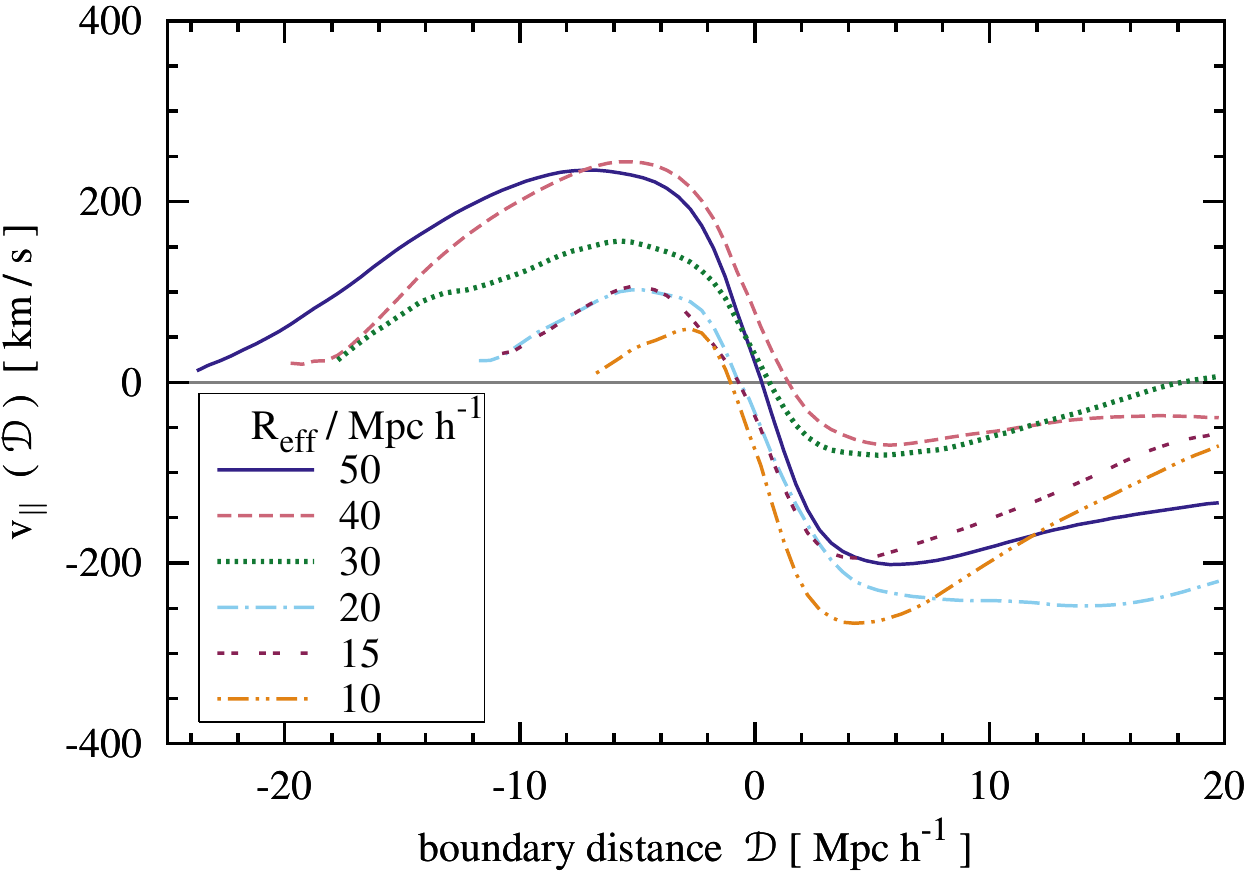}
    \end{array}$
     \caption{ The peculiar velocity profile of the six randomly selected voids shown in \reffig{fig:density_profiles_random_voids}. It shows the velocity component, $\vrad$, along the direction of $\Vector{r}$ and $\distVec{}$, respectively. The two panels show the spherical (top) and \boundary{} (bottom) profiles of those voids. }
     \label{fig:velocity_profiles_random_voids}
\end{figure}

\MCn{For investigating void velocity profiles we use the same objects, both individual and stacked samples of voids, as we used when studying the density profiles in \refsec{sec:density_profile}. \reffigS{fig:velocity_profiles_random_voids}{fig:velocity_profiles_stacked} show the corresponding $\vrad$ profiles for individual and stacked voids. For brevity, we focus our discussion on the stacked velocity profiles, with individual voids showing similar trends, albeit with large individual variations.} 

\begin{figure}
     \centering
    $\begin{array}{c}
        \includegraphics[width=\linewidth,angle=0]{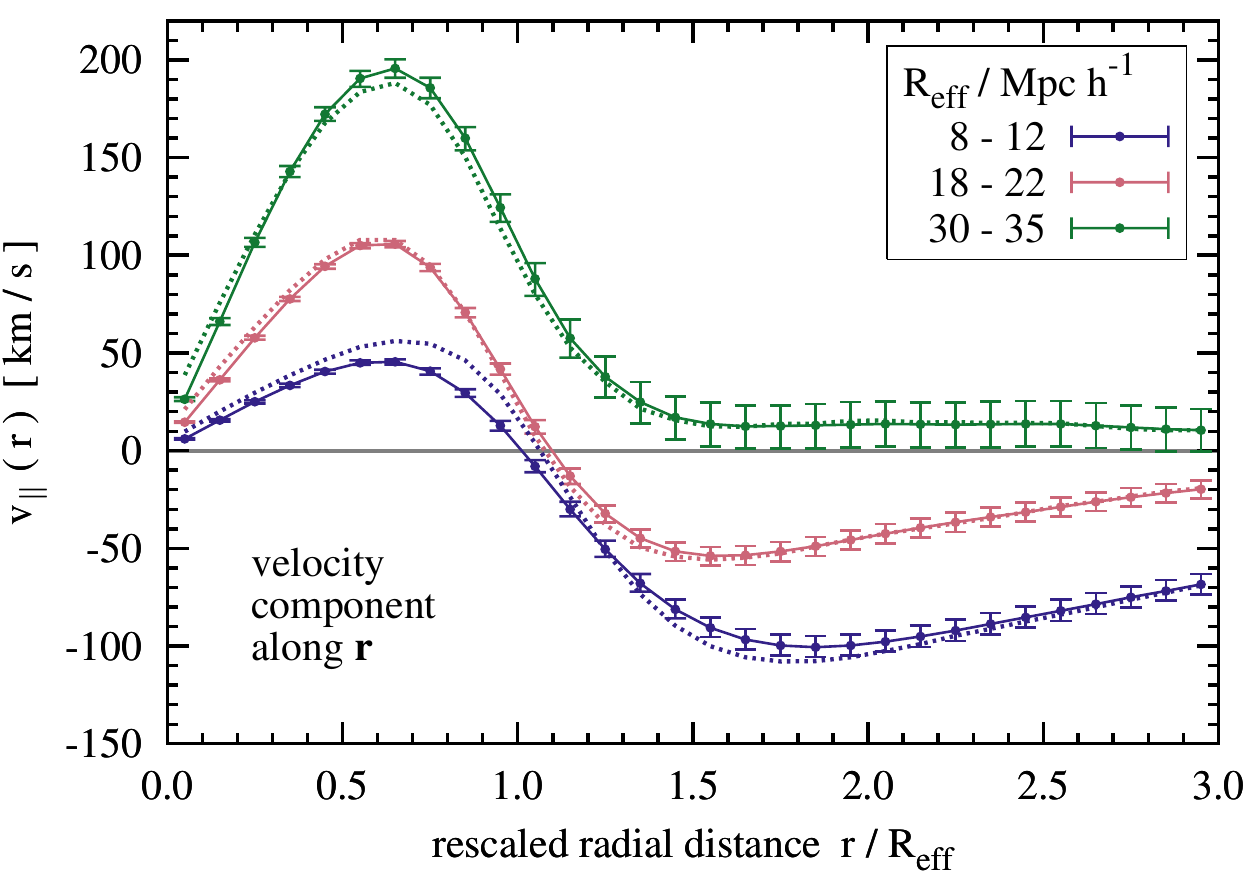} \\
        \includegraphics[width=\linewidth,angle=0]{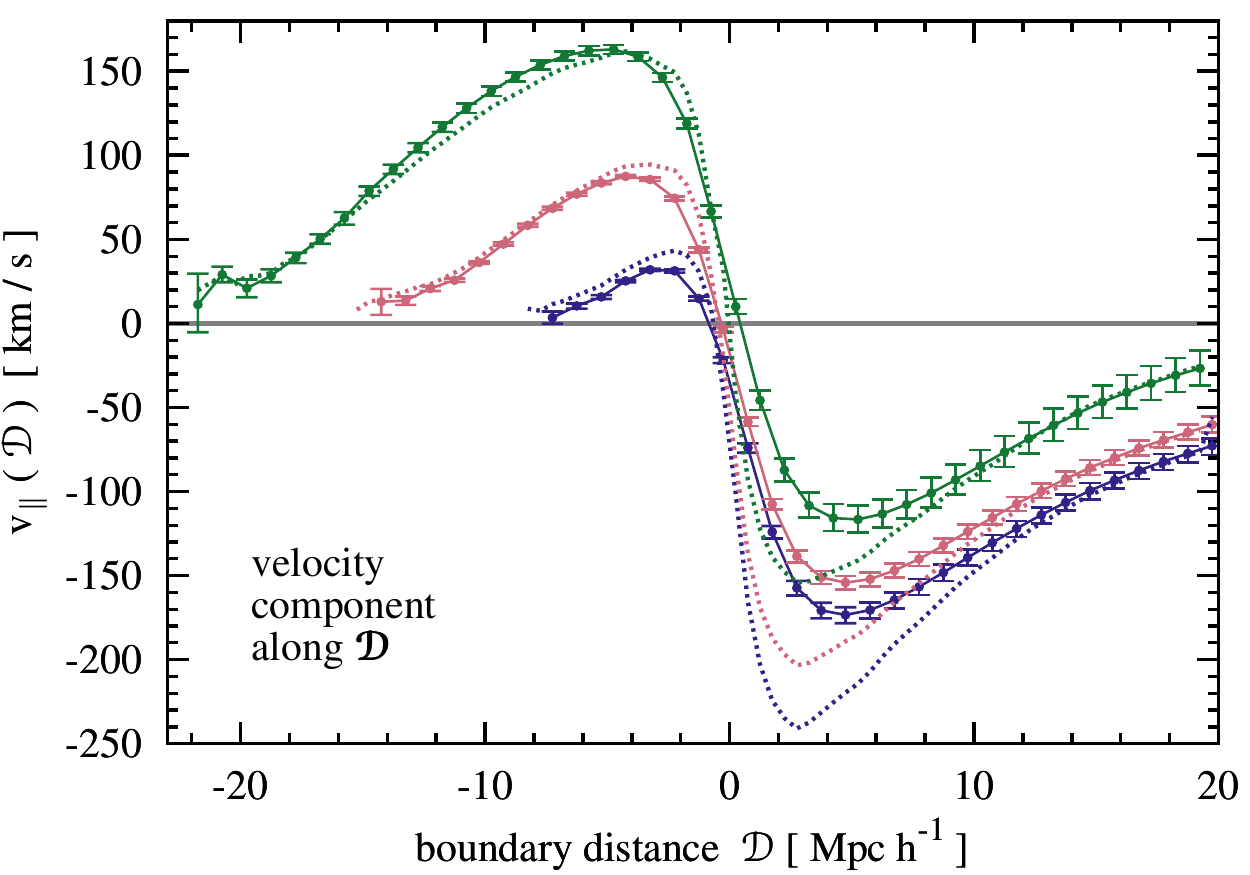}
    \end{array}$
     \caption{ The peculiar velocity profile as a function of radial distance, $r$, for spherical stacking (top) and as a function of void boundary distance, $\dist$, for \boundary{} stacking (bottom). It shows the velocity component, $\vrad$, along the direction of $\Vector{r}$ and $\distVec{}$, respectively. To guide the eye, the data points are connected with solid lines. The dotted lines show the linear theory prediction for $\vrad$ given the average density profiles of \reffig{fig:density_profiles_stacked}. }
     \label{fig:velocity_profiles_stacked}
\end{figure}

The spherical velocity profiles show outflows from voids, which peak at  ${\sim}0.6\reff$, and that are followed by regions with lower outflow velocities or even inflows. The nearly-linear increasing outflow for $r{\lsim}0.5\reff$ indicates that void interiors expand faster than the average universe showing a so-called super-Hubble outflow \citep{Icke1984,vandeWeygaert2011,Aragon-Calvo2013}.
For the \emph{\boundary{}} profile, the velocity, $\vrad$, increases until near the void boundary, $\dist\lsim-3\Mpch$, and is then followed by a rapid switch from outflow to inflow. This behaviour at $\dist\simeq0$ is consistent with infall onto the void boundary, which, given its high density, is the main local driver of dynamics. At further distances from the void boundary, the velocity slowly converges towards $0$, as expected.

Given the density profiles shown in \reffig{fig:density_profiles_stacked}, we can use linear theory to predict the $\vrad$ values \citep[e.g. see][]{vandeWeygaert1993}, which are shown as dotted lines in \reffig{fig:velocity_profiles_stacked}. The linear predictions are given by
\begin{equation}
    v_{\rm \parallel, lin} = -\frac{H f}{\overline{\rho}_{\rm m}} \frac{M(<x)}{S(x)}
    \label{eq:velocity_linear_approximation} \;,
\end{equation}
with $H$ the Hubble factor, $f\simeq\Omega_{\rm m}^{0.55}$ the linear growth factor and $\overline{\rho}_{\rm m}$ the mean background density of matter. The symbol $x$ stands for the radial distance, $r$, for spherical profiles and for the distance, $\dist{}$, for \boundary{} profiles. The factor $M(<x)$ denotes the mass contrast inside $x$ and $S(x)$ denotes the area of a surface of constant $x$. See \refappendix{appendix:velocity_linear} for details and for a short derivation of the relation.

The linear theory prediction agrees with the data for the spherical profile, except for a few small systematic effects: the velocity of small voids is overpredicted while that of large voids is underpredicted. These discrepancies, seen also by \citet{Hamaus2014b}, have been attributed to the effect of surrounding structures on void interiors \citep{Ruiz2015}. In the case of \boundary{} profiles, the linear theory is in agreement only for the void interior, i.e. $\dist\lsim0$, and at large distances, $\dist\gsim10\Mpch$. Large discrepancies are present at the void boundary and just outside the void where the linear predictions can be off by up to $100\kms$. Such differences are not surprising since linear theory is valid in the regime $|\delta|\ll1$. For spherical \MCn{stacking}, while the average $\delta$ is not very small, it is below unity at every point. In contrast, the \boundary{} \MCn{stacking} has very large values of $\delta$, as high as 3 at the void edge, which explains why large discrepancies are seen only at, and just outside, the void boundary. \MCn{Moreover, as shown in \reffig{fig:density_profiles_random_voids}, individual voids have density values above unity for spherical profiles as well, so linear theory would break down in such cases too. The difference is that for spherical profiles the position of the $\delta>1$ region varies from void to void, so departures from linear theory average out when stacking many such objects, whereas for the \boundary{} profile the departures are always at the same position, $\dist\simeq0$.}

\begin{figure}
     \centering
     \includegraphics[width=\linewidth,angle=0]{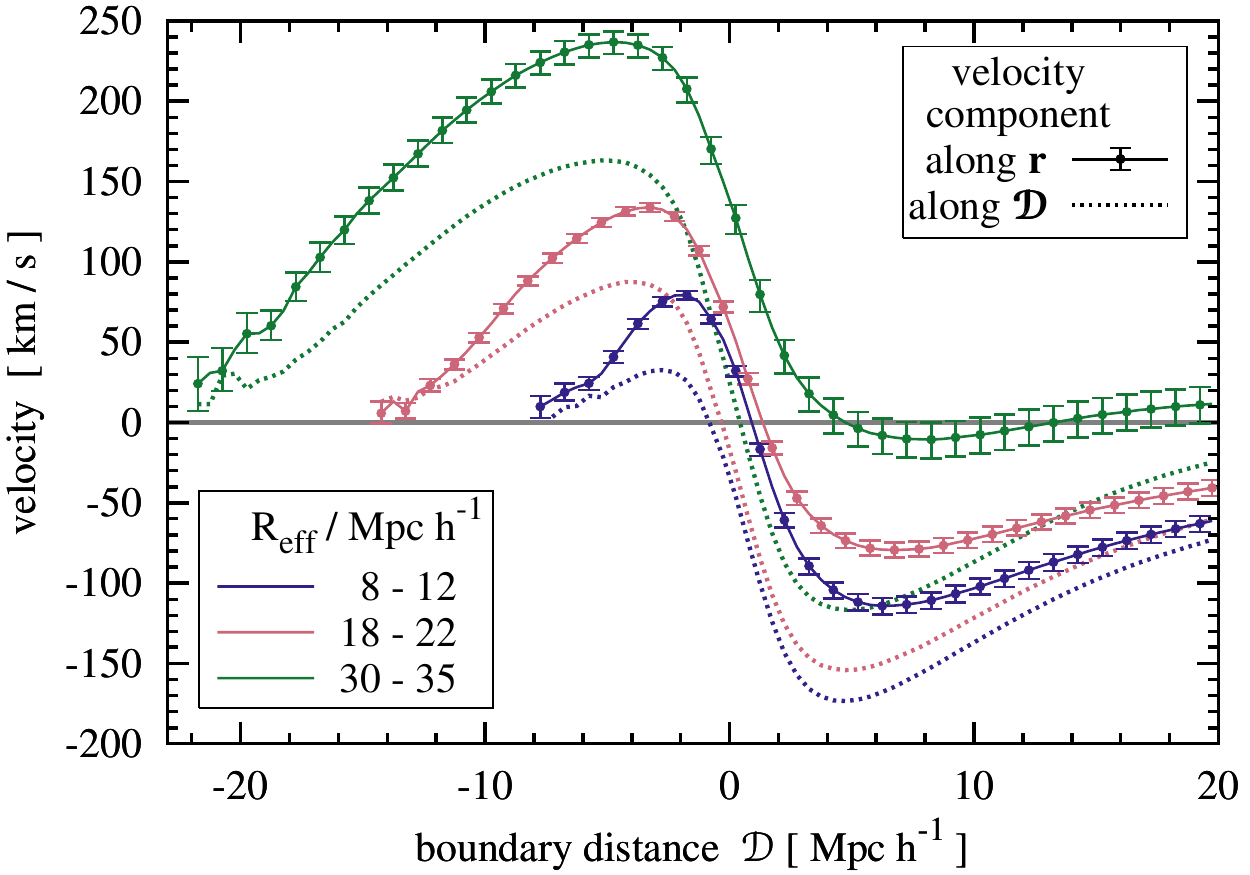} 
     \caption{ \MCn{The boundary profiles of the velocity component along the radial vector, $\Vector{r}$, (solid with symbols) and along the boundary distance vector, $\distVec$ (dotted). It shows that the outflow from voids is preferentially along the radial direction increasing until close to the void edge.} }
     \label{fig:velocity_comparison}
\end{figure}

\MCn{We find that the $\vrad{}$ peak is highest for spherical profiles and that the same peak is up to $20\%$ lower for \boundary{} profiles, even though in the latter case the velocity increases until close to the void edge. To explain this, \reffig{fig:velocity_comparison} shows the \boundary{} profile for the velocity component along the radial direction. For comparison, the dashed lines show the profile of the velocity component along $\distVec$, which corresponds to the solid lines with symbols in the bottom panel of \reffig{fig:velocity_profiles_stacked}. For $\dist<0$, the radial velocity is larger than the velocity component along $\distVec$ which shows that the outflow from voids is preferentially directed radially. \reffig{fig:velocity_comparison} also shows that the radial velocity, when binned according to $\dist$, increases until close to the void boundary and it is then followed by a very steep decrease at the edge of the void. This contrasts with the spherical profile of the radial velocity (see top panel of \reffig{fig:velocity_profiles_stacked}), which shows a peak at $r\sim0.6\reff{}$ and not at the edge of the void, i.e. $r\sim\reff{}$.}

\begin{figure}
     \centering
    $\begin{array}{c}
        \includegraphics[width=\linewidth,angle=0]{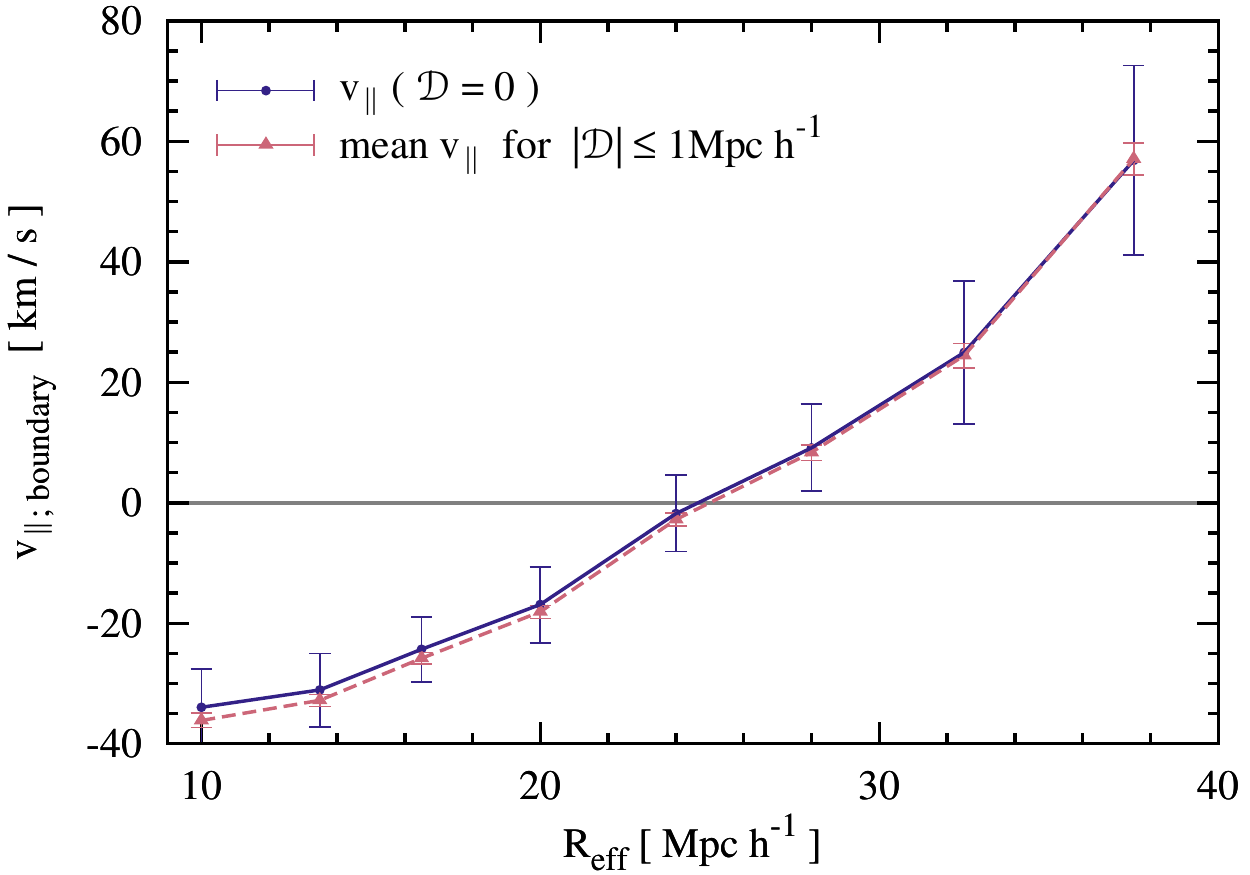} \\
        \includegraphics[width=\linewidth,angle=0]{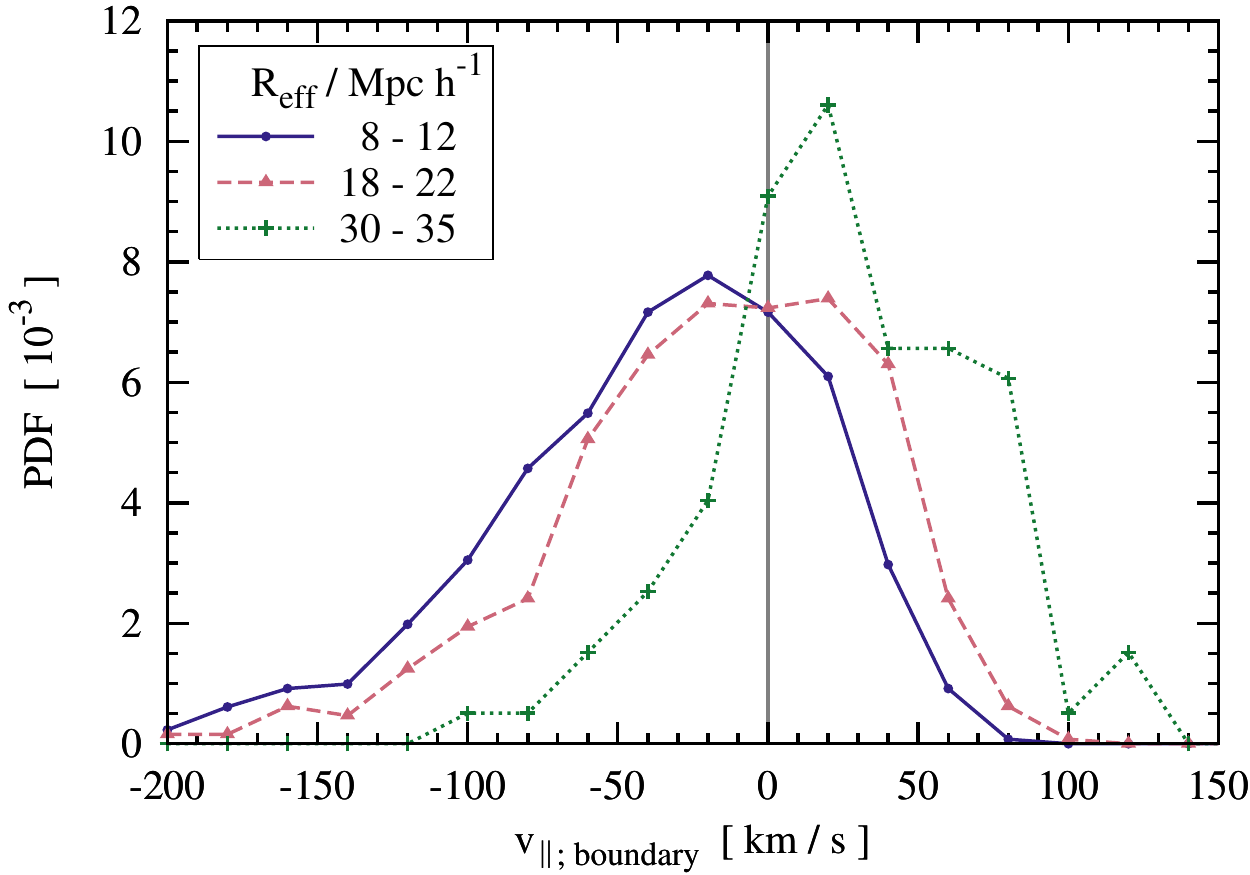}
    \end{array}$
     \caption{ The velocity of the void boundary, $v_{\rm \parallel;\; boundary}$, as a function of void size. Negative values correspond to contracting voids and positive values to expanding voids. The top panel shows this velocity for voids stacked according to their size, $\reff{}$. It shows the velocity at $\dist=0$ (solid curve) and the mean velocity over the interval $|\dist|\le1\Mpch$ (dashed curve), which is more robust. The bottom panel shows the probability distribution function (PDF) of the ridge velocity for individual voids of various sizes. The distribution is very broad with each sample having both expanding and contracting voids. }
     \label{fig:velocity_boundary}
\end{figure}

The \boundary{} profile offers a natural way of discriminating between expanding and contracting voids. For example, expanding voids correspond to a positive $\vrad{}$ value at their boundary since the boundary is moving outwards. The top panel of \reffig{fig:velocity_boundary} shows the values of the velocity at the boundary, $\dist=0$, and also the $\vrad{}$ value averaged over the interval $|\dist{}|\le1\Mpch$, with the latter being less prone to noise. The plot shows that, on average, small voids are contracting while large ones are expanding, with voids of $\reff\sim25\Mpch$ being at the transition between the two behaviours. \MCnnn{Using the mean density of the large-scale region in which voids are embedded, previous studies have characterised the voids as under- or overcompensated \citep[e.g.][]{Ceccarelli2013,Hamaus2014b,Nadathur2015c}. For example, the smallest two void stacks in \reffig{fig:velocity_profiles_stacked} are overcompensated while the larger voids are slightly undercompensated. This distinction can be determined, for example, using the sign of the radial velocity at $r\gsim1.5\reff{}$ (see top panel of \reffig{fig:velocity_profiles_stacked}), with positive values corresponding to underdense regions and vice versa. Combining this with our analysis of the void boundary dynamics, we find that overcompensated voids are predominantly contracting while the undercompensated ones are predominantly expanding.}

Using the \boundary{} profile one can determine even for individual voids if they are expanding or contracting, as we show in the bottom panel of \reffig{fig:velocity_boundary}. For example, while most small voids are contracting, there is also a significant fraction that are expanding. Similarly for the largest voids: while most are expanding, there are large contracting voids too. \MCn{Thus, expanding and contracting voids cannot be differentiated using just their size, $\reff{}$, and additional void properties need to be considered \citep[see e.g.][]{Nadathur2015c}.}

%%%%%%%%%%%%%%%%%%%%%%%%%%%%%%%%%%%%%%%%%%%%%%%%%%%%%%%%%%%%%%%%%%%%%%%%%%%%%%
\section{Weak lensing from voids}
\label{sec:weak_lensing}
We now address how \boundary{} stacking can be used to enhance the weak lensing signal of voids. Since it is a small effect, void lensing is difficult to measure \citep{Melchior2014}, although recently multiple detections of this signal have been reported \citep{Clampitt2014,Gruen2015}. Increasing the signal to noise of this measurement, by either having a larger sample of voids and/or by improving how voids are stacked, would result in a powerful cosmological probe, especially for tests of modified gravity theories \citep{Cai2014,Barreira2015}.

\begin{figure}
     \centering
    $\begin{array}{c}
        \includegraphics[width=\linewidth,angle=0]{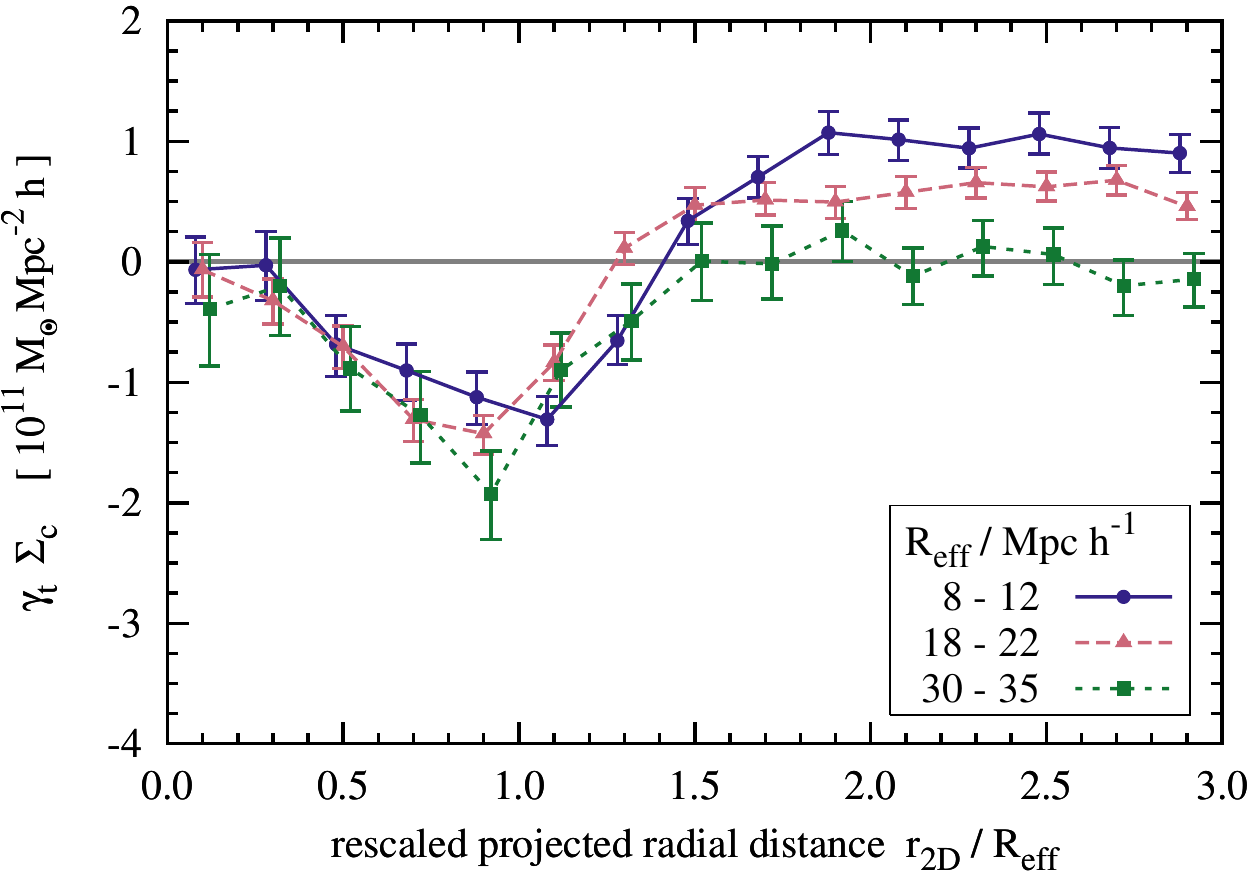} \\
        \includegraphics[width=\linewidth,angle=0]{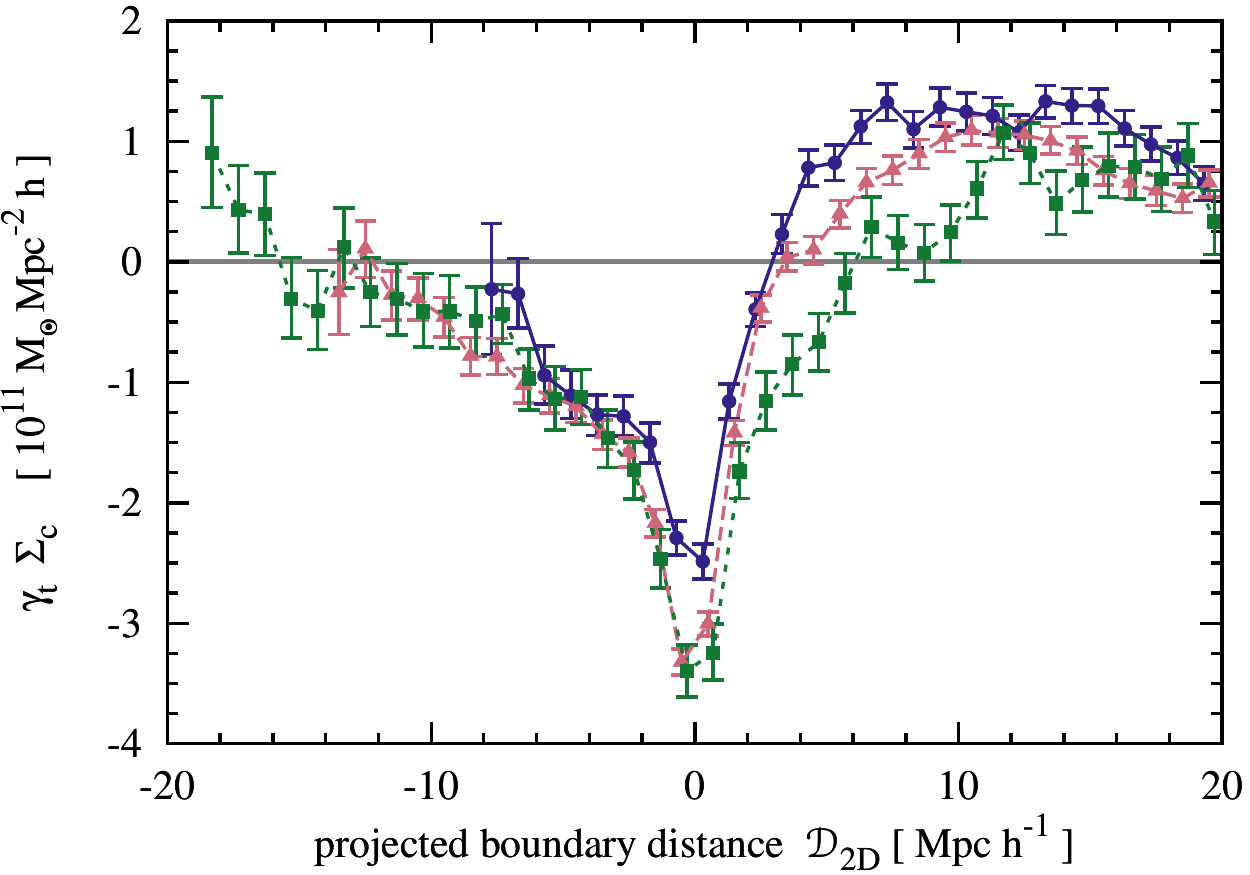}
    \end{array}$
     \caption{ The stacked tangential shear, $\shear{}$, of voids in three ranges in effective radius, $\reff{}$. The top panel shows the spherically averaged result. The bottom panel shows the result when voids are stacked with respect to their boundary. The error bars show the 1$\sigma$ uncertainties due to object-to-object variation. }
     \label{fig:lensing_tangential_shear}
\end{figure}

\begin{figure}
     \centering
    $\begin{array}{c}
        \includegraphics[width=\linewidth,angle=0]{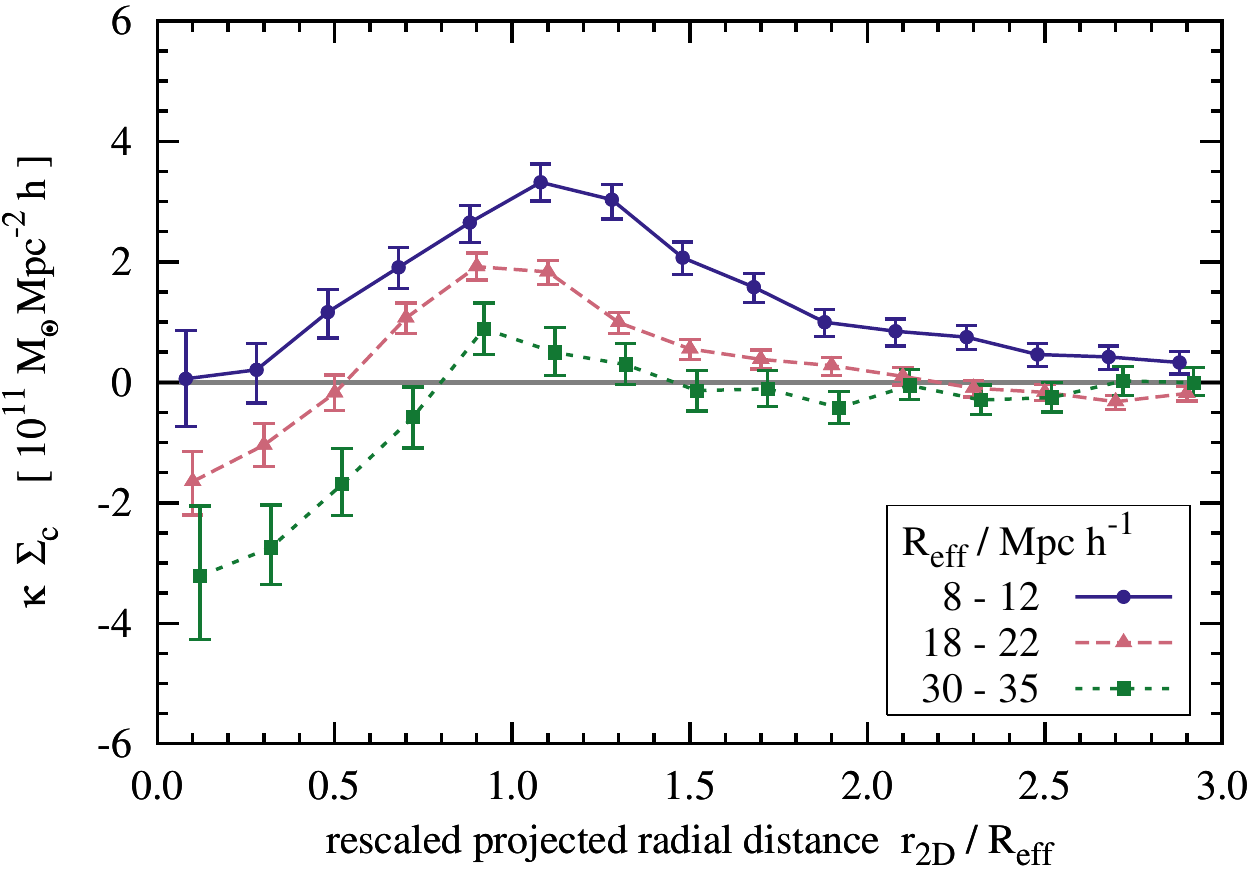} \\
        \includegraphics[width=\linewidth,angle=0]{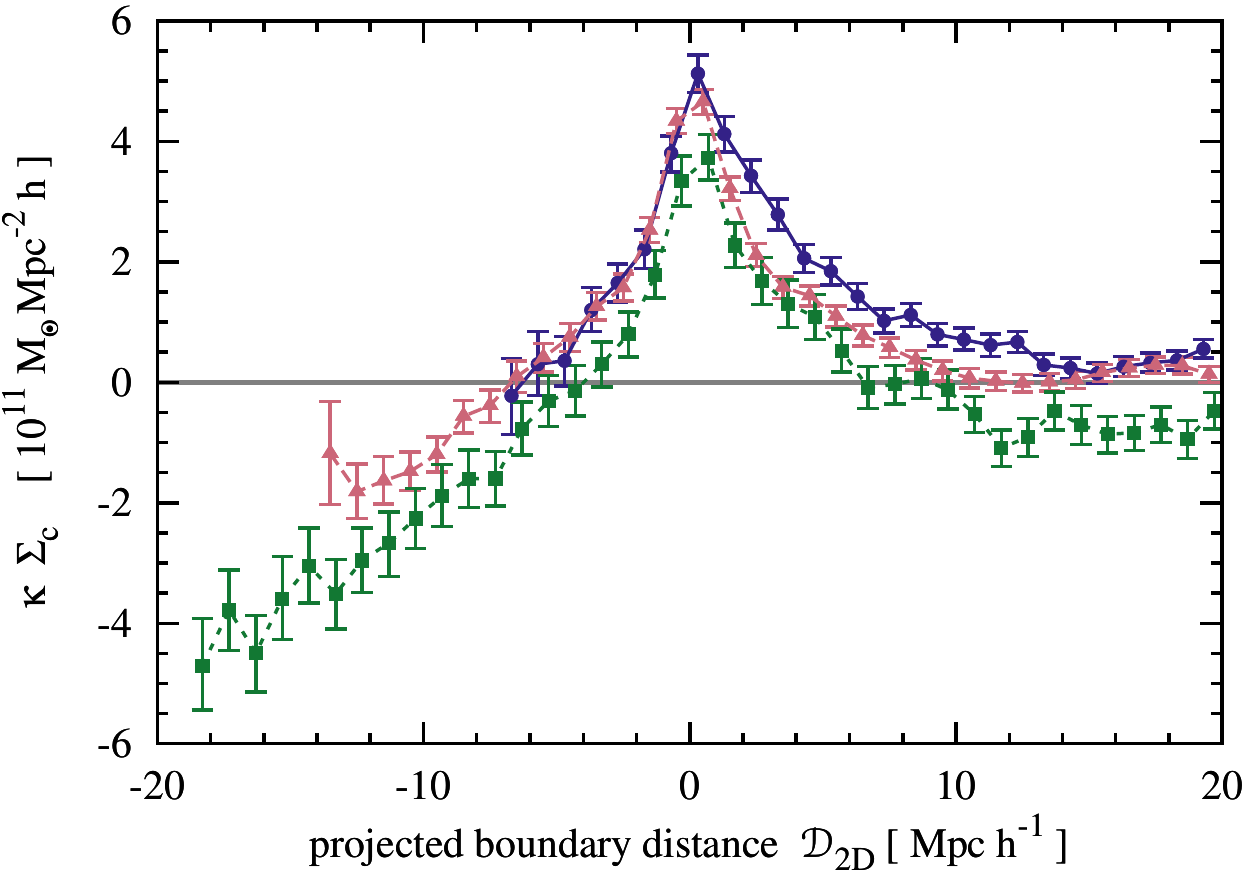}
    \end{array}$
     \caption{ Same as \reffig{fig:lensing_tangential_shear}, but for the the stacked lensing convergence, $\conv{}$, of voids. }
     \label{fig:lensing_convergence}
\end{figure}

\MCn{Within the thin lens and the Born approximation, the weak lensing signal is determined by the surface mass density,}
\begin{equation}
    \Sigma(\bm{\xi}) = {\overline{\rho}_m} \int \delta(\bm{\xi},z) \mathrm{d}z
    \label{eq:lensing_smd} \;,
\end{equation}
where $\bm{\xi}$ is the position vector in the plane of the lens and $z$ is the direction along the line-of-sight. We compute $\Sigma(\bm{\xi})$ for three lines-of-sight that correspond to the simulation principal axes. For each line-of-sight we obtain $\Sigma(\bm{\xi})$ on a $1280^2$ regular grid with grid spacing $0.39\Mpch$. \MCn{We then proceed to compute the lensing potential, $\Psi$, via the relation 
\begin{equation}
     \nabla^2_{\xi}\Psi(\bm{\xi}) =2\frac{\Sigma(\bm{\xi}) }{\Sigma_{\rm c}}
     \; ,
\end{equation}
with the Laplacian operator restricted to the plane of the lens.} The  quantity, $ \Sigma_c = c^2D_{\rm S}/{(4\pi G D_{\rm L}D_{\rm LS})} $, is the critical surface mass density for lensing, where $D_{\rm S}$, $D_{\rm L}$ and $D_{\rm LS}$ denote the angular diameter distance between the observer and the source, the observer and the lens, and the lens and the source. The exact value of $\Sigma_c$, which depends on the characteristics of the lensing survey, is unimportant when comparing between the spherical and boundary stacking approaches. 

\MCn{For each point, we compute the convergence, $\kappa$, and the shear, ${\bm\gamma}=(\gamma_1,\gamma_2)$, as
\begin{align}
    \kappa(\bm{\xi})   & = \tfrac{1}{2} \left[\Psi_{11}(\bm{\xi}) + \Psi_{22}(\bm{\xi})\right]  \equiv \Sigma(\bm{\xi}) / \Sigma_{\rm c}\\
    \gamma_1(\bm{\xi}) & = \tfrac{1}{2} \left[\Psi_{11}(\bm{\xi}) - \Psi_{22}(\bm{\xi})\right] \\
    \gamma_2(\bm{\xi}) & = \Psi_{12}(\bm{\xi}) \equiv \Psi_{21}(\bm{\xi})
    \label{eq:lensing_shear} \;,
\end{align}
where the subscripts of $\Psi$ denote derivatives with respect to the two coordinate axes in the plane of the lens.}

\MCn{For \emph{spherical} stacking, the lensing signal is averaged as a function of the projected radial distance, $r_{\rm 2D}$, from the void centre. This results in the convergence, $\kappa(r_{\rm 2D})$, which is a mean value inside a spherical shell of radius $r_{\rm 2D}$. In the case of the shear, we are interested in the tangential component, $\gamma_{\rm t}$, given by
\begin{align}
    \gamma_{\rm t} & =  -\gamma_1 \cos(2\theta) + \gamma_2 \sin(2\theta)
    \label{eq:lensing_tangential_shear} \;,
\end{align}
where $\theta$ is the angle between the first coordinate axis and the position of the point with respect to the void centre. After computing $\kappa(r_{\rm 2D})$ and $\gamma_{\rm t}(r_{\rm 2D})$ for each void, we stack the voids according to their effective radius and across the three different lines-of-sight for which we computed $\Sigma$. Since the projected matter distribution is different along those orthogonal lines-of-sight, averaging their lensing signal increases the signal-to-noise ratio.}

For \emph{\boundary{}} stacking, the procedure is slightly different, since we need to identify the boundary of the void in the lens plane. We do so by slicing the boundary of the void, which is a 2D surface, along the plane of the lens, with the slice centred at the point inside the void that is the farthest from the void boundary (this is the point corresponding to the minimum distance, $\dist_{\rm min}$). Following this, we obtain a closed curve in the lens plane that corresponds to one particular choice of the void boundary (see discussion below), which is then used to compute the distance in the plane of the lens, $\dist_{\rm 2D}$, of each surface element. Following this, for every void we compute the mean value of the convergence as a function of $\dist_{\rm 2D}$ resulting in the quantity $\kappa(\dist_{\rm 2D})$. \MCn{The tangential shear is computed using \eq{eq:lensing_tangential_shear} but with $\theta$ denoting the angle between the first coordinate axis and the 2D boundary distance vector, $\distVec{}_{\rm 2D}$, at that point. Finally, we stack all voids of similar size and across the three lines-of-sight.}

We note that this is just one possible choice for stacking with respect to the void \boundary{}, and may not be the optimal choice. For lensing studies, it is better to identify 2D voids in thin redshift slices, since this greatly enhances the lensing signal \citep{Clampitt2014}. The boundary of these 2D voids is a 1D curve in the plane of the sky. In such a case there is no ambiguity in choosing the 1D void boundary in the plane of the lens.

In \reffig{fig:lensing_tangential_shear} we show the void tangential shear obtained using the two stacking procedures. The spherically averaged $\shear$ shows the characteristic dip of void lensing at $r_{\rm 2D}\simeq\reff{}$, which is nearly the same for the three void samples. This depression is more pronounced when using \boundary{} stacking for which the signal is twice as large. Using \boundary{} stacking increases the convergence, $\conv$, also by a factor of about two, as can be inferred from \reffig{fig:lensing_convergence}. This doubling of the lensing signal is the result of a better separation between the void border, where most of the mass is, and the void interior, which is mostly empty. This factor of two represents only a lower limit to the potential improvements resulting from the use of \boundary{} stacking. Likely, the gain can be increased further by optimizing the selection of the void boundary in the plane of the sky. 

%The tangential shear 
%A back of the envelope calculation, using that \boundary{} density profiles show a $\Delta\delta=4$ change in overdensity between the interior and boundary of the void, compared with only a $\Delta\delta=1$ increase for spherical profiles, suggests a potential gain as high as a factor of four. 

%%%%%%%%%%%%%%%%%%%%%%%%%%%%%%%%%%%%%%%%%%%%%%%%%%%%%%%%%%%%%%%%%%%%%%%%%%%%%%%%%%%%%%%%%
\section{Discussion and conclusions}
\label{sec:conclusions}
We have proposed a new method for characterising voids that has several advantages over the conventional spherical approach, as demonstrated by our analysis of galaxy voids in the Millennium cosmological simulation. This approach, which we call the \emph{\boundary{} profile}, is based on describing the structure of voids as a function of the distance from their boundary, which allows for a natural segregation of the inner, boundary and outer regions of each void. 

Voids are characterised by two defining features: they consist of large, fairly underdense volumes, with the evacuated matter found in a thin overdense region at the boundary, and they have very complex, non-spherical, shapes. The spherical averaging approach is inadequate for describing voids due to this very combination of features, as we exemplify for a simplified void model (\reffigS{fig:toy_example}{fig:toy_example_profiles}) and for realistic voids (\reffigS{fig:density_profiles_random_voids}{fig:density_profiles_stacked}). This is a consequence of the fact that taking a spherical average over an intrinsically non-spherical object leads to a complex juxtaposition of the inner, border and outer regions of that object, with each region having very different density. By contrast, the \boundary{} profile method differentiates, by construction, between those regions.

The \boundary{} profile analysis revealed that the interior of voids is characterised by low densities that increase slowly towards the void boundary. This is followed by a steep rise of a density ridge at the void boundary, which decreases nearly as fast outside the void. The peak of the density ridge corresponds to $1+\delta\simeq4$ while the interior of the void has $1+\delta\simeq0.2-0.4$. \MCn{We found a simple fitting function (Eq. \ref{eq:fit_function_density}) that describes fairly well the void density profiles and that can be parametrized in terms of a single quantity, the void effective radius (see \reffig{fig:best-fitting_parameters}). This parametrization provides a convenient way of describing the variation of density profiles with void size and allows for simple comparisons to theoretical models of void evolution, such as the spherical top-hat underdensity model (\SvdW{}).} 

The \boundary{} density profile is self-similar, i.e. independent of void size, after rescaling the distance coordinate by the thickness of the void's inner density ridge (see \reffig{fig:density_profiles_scaling}). This suggests that the void interior knows about the void boundary or vice versa, and that the evolution of the two is coupled. This simple behaviour is reminiscent of the self-similar nature of dark matter haloes \citep{Navarro1996,Navarro1997} whose origin, while not well understood, must reflect the scale-free nature of gravity. \MCn{In contrast to haloes for which the characteristic scale is determined by the matter distribution in the innermost region, for voids the characteristic scale is determined by the matter distribution at the edge of the void.}

The \boundary{} profile of the peculiar velocity reveals outflows from voids, which peak at a few Megaparsecs from the edge of the void, and an external infall region onto the void boundary. \MCn{These outflows are preferentially directed along the radial direction, with the radial velocity being larger than the velocity component pointing towards the closest void edge (see \reffigS{fig:velocity_profiles_stacked}{fig:velocity_comparison}).} The \boundary{} profiles are especially suited for capturing the infall onto the void boundary, which is not seen for spherical profiles, and for determining if the voids are contracting or expanding (see \reffig{fig:velocity_boundary}).

The \boundary{} stacking method increases the weak lensing signal of voids by at least a factor of two when compared to the classical spherical stacking method. This gain can potentially be further increased by optimizing the selection of the void boundary on the plane of the sky (Cautun et al., in prep.). This gain in lensing signal boosts the utility of voids as cosmological probes, especially when applied to future large volume surveys like DESI \citep{Levi2013}, LSST \citep{LSST2009} and Euclid \citep{Laureijs2011}.

%%%%%%%%%%%%%%%%%%%%%%%%%%%%%%%%%%%%%%%%%%%%%%%%%%%%%%%%%%%%%%%%%%%%%%%%%%%%%%%%%%%%%%%%%%
\section*{Acknowledgements}
We thank Rien van de Weygaert for insightful and valuable comments on an earlier draft of the paper and we are grateful to Alexandre Barreira, Shaun Cole, Jiaxin Han, Baojiu Li and Nuala McCullagh for helpful discussions.
We also thank the anonymous referee for their comments that have helped us improve the paper.
This work was supported in part by ERC Advanced Investigator grant COSMIWAY 
[grant number GA 267291] and the Science and Technology Facilities Council 
[grant number ST/F001166/1, ST/I00162X/1]. YC was supported by the Durham Junior Research Fellowship

This work used the DiRAC Data Centric system at Durham University, 
operated by ICC on behalf of the STFC DiRAC HPC Facility (www.dirac.ac.uk). 
This equipment was funded by BIS National E-infrastructure capital 
grant ST/K00042X/1, STFC capital grant ST/H008519/1, and STFC DiRAC 
Operations grant ST/K003267/1 and Durham University. DiRAC is part 
of the National E-Infrastructure. Data from the Millennium/Millennium-II 
simulation is available on a relational database accessible from 
http://galaxy-catalogue.dur.ac.uk:8080/Millennium .

%%%%%%%%%%%%%%%%%%%%%%%%%%%%%%%%%%%%%%%%%%%%%%%%%%%%%%%%%%%%%%%%%%%%%%%%%%%%%%%%%%%%%%%%%

%\newcommand{\jcap}{JCAP}
%{Journal of Cosmology and Astroparticle Physics}
\bibliographystyle{mnras}
\bibliography{void_reference}

\appendix

%%%%%%%%%%%%%%%%%%%%%%%%%%%%%%%%%%%%%%%%%%%%%%%%%%%%%%%%%%%%%%%%%%%%%%%%%%%%%%%%%%%%%%%%%
\section{Velocity field in the linear approximation}
\label{appendix:velocity_linear}
In the linear approximation, the peculiar velocity at redshift, $z=0$, is given by \citep{Peebles1980}
\begin{equation}
    \Vector{v} = \frac{Hf}{4\pi G\overline{\rho}_{\rm m}} \Vector{g}
    \label{eq:peculiar_velocity_general}\; ,
\end{equation}
where $G$ is the gravitational constant and $\Vector{g}$ is the gravitational field (for the remaining symbols see Eq. \ref{eq:velocity_linear_approximation}). The same relation holds for the velocity component, $\vrad{}$, along either $\Vector{r}$ or $\distVec{}$, but with $\Vector{g}$ replaced by $\grad{}$. %This equation has been derived under two assumptions. Firstly, by discarding the $\Vector{\nabla}\cdot(\delta\Vector{v})$ term in the continuity equation, which is valid in the limit $\Vector{\nabla}\cdot(\delta\Vector{v})\ll \Vector{\nabla}\cdot\Vector{v}$, i.e. $\delta\ll1$. Secondly, 

\begin{figure}
     \centering
     \includegraphics[width=\linewidth,angle=0]{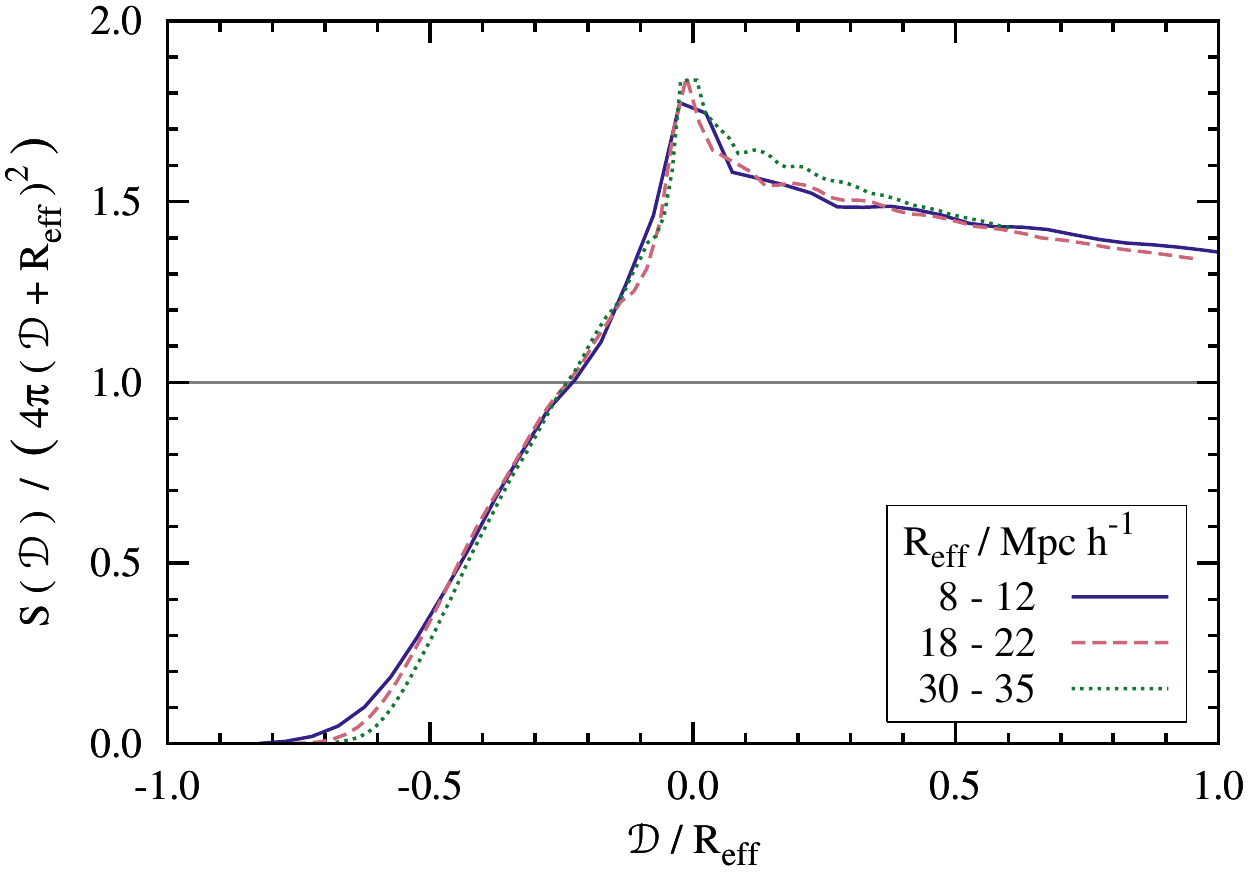} 
     \caption{ The area of a surface, $S(\dist)$, of constant void boundary distance, $\dist{}$. It shows the mean area $S(\dist)$, normalized by the area of a sphere of radius, $\dist{}+\reff$, as a function of the normalized void boundary distance, $\dist{}/\reff$. The three curves give the average value over all the voids in their respective $\reff$ intervals. }
     \label{fig:surface_area}
\end{figure}

Applying Gauss' theorem to the gravitational field, we have
\begin{equation}
    \int_{\rm S} \Vector{g} \cdot \mathrm{d} S(x) = -4\pi G M(<x)
    \label{eq:Gauss_law}\; ,
\end{equation}
where $x$ stands for either $r$ or $\dist{}$ and, 
\begin{equation}
    M(<x) = \int_{x_{\rm min}}^{x} \delta(x') S(x') \mathrm{d} x'
    \label{eq:mass_contrast} \; ,
\end{equation}
is the mass contrast enclosed by the surface, $S(x)$, of constant $x$ values. In the case of $x=r$, the lower integration bound, $x_{\rm min}=0$, and the surface, $S$, corresponds to a spherical surface. For $x=\dist{}$, $x_{\rm min}$ gives the distance, $\dist_{\rm min}$, from the boundary of the farthest point inside the void (see \refsec{subsec:data_shape_profiles}), while the surface, $S$, has an irregular shape. \reffig{fig:toy_example} shows a cross section through an example void. \eq{eq:Gauss_law} can be rewritten as,
\begin{equation}
    \overline{\grad} = - \frac{4\pi G M(<x)} {S(x)}
    \label{eq:mean_g_parallel} \; ,
\end{equation}
where $\overline{\grad}$ denotes the average value of $\grad{}$ over the surface $S(x)$. Inserting this last expression into \eq{eq:peculiar_velocity_general} results in \eq{eq:velocity_linear_approximation} used to compute $\vradlin(r)$ and $\vradlin(\dist)$. \MCn{Note that since \eq{eq:velocity_linear_approximation} is not linear in $S(x)$, one needs to compute the linear theory predictions separately for each void, using their own density profile, and only in the final step to average over all the voids in the stack.} 

To compute $\vradlin(\dist)$ one needs to know the function $S(\dist{})$. This depends on the shape of the void boundary and, due to the large diversity of watershed void shapes, is different for each void. \reffig{fig:surface_area} shows the mean value of $S(\dist{})$, as measured for \MI{} voids of various sizes. It shows that, when scaled appropriately, $S(\dist{})$ is approximately independent of the effective void radius. The scaled $S(\dist{})$ is maximal for $\dist{}=0$ since the void boundary is the most irregular $\dist{}=\text{constant}$ surface, as may be appreciated from \reffig{fig:toy_example}. In the limit, $\dist\gg\reff$, the surface $S(\dist)$ becomes a sphere and hence the scaled area shown in \reffig{fig:surface_area} converges to 1.

\end{document}